\documentclass[sigconf,screen]{acmart}
\AtBeginDocument{%
  \providecommand\BibTeX{{%
    \normalfont B\kern-0.5em{\scshape i\kern-0.25em b}\kern-0.8em\TeX}}}

\newcommand{\streamered}{LifeStream}
\newcommand{\streamer}{LifeStream }
\newcommand{\DOI}{\href{https://doi.org/10.5281/zenodo.4331660}{https://doi.org/10.5281/zenodo.4331660 }}
\newcommand{\github}{\href{https://github.com/anandj91/LifeStream}{https://github.com/anandj91/LifeStream }}

\usepackage[draft]{minted}
\usepackage[normalem]{ulem}
\usepackage{todonotes}
\usepackage{listings}
\usepackage{caption}
\usepackage{subfigure}
\usepackage{tabularx}
\usepackage[ruled,vlined]{algorithm2e}
\usepackage{hyperref}
\usepackage{microtype}
\usepackage{enumitem}

\makeatletter
\def\PYGdefault@reset{\let\PYGdefault@it=\relax \let\PYGdefault@bf=\relax%
    \let\PYGdefault@ul=\relax \let\PYGdefault@tc=\relax%
    \let\PYGdefault@bc=\relax \let\PYGdefault@ff=\relax}
\def\PYGdefault@tok#1{\csname PYGdefault@tok@#1\endcsname}
\def\PYGdefault@toks#1+{\ifx\relax#1\empty\else%
    \PYGdefault@tok{#1}\expandafter\PYGdefault@toks\fi}
\def\PYGdefault@do#1{\PYGdefault@bc{\PYGdefault@tc{\PYGdefault@ul{%
    \PYGdefault@it{\PYGdefault@bf{\PYGdefault@ff{#1}}}}}}}
\def\PYGdefault#1#2{\PYGdefault@reset\PYGdefault@toks#1+\relax+\PYGdefault@do{#2}}

\expandafter\def\csname PYGdefault@tok@w\endcsname{\def\PYGdefault@tc##1{\textcolor[rgb]{0.73,0.73,0.73}{##1}}}
\expandafter\def\csname PYGdefault@tok@c\endcsname{\let\PYGdefault@it=\textit\def\PYGdefault@tc##1{\textcolor[rgb]{0.25,0.50,0.50}{##1}}}
\expandafter\def\csname PYGdefault@tok@cp\endcsname{\def\PYGdefault@tc##1{\textcolor[rgb]{0.74,0.48,0.00}{##1}}}
\expandafter\def\csname PYGdefault@tok@k\endcsname{\let\PYGdefault@bf=\textbf\def\PYGdefault@tc##1{\textcolor[rgb]{0.00,0.50,0.00}{##1}}}
\expandafter\def\csname PYGdefault@tok@kp\endcsname{\def\PYGdefault@tc##1{\textcolor[rgb]{0.00,0.50,0.00}{##1}}}
\expandafter\def\csname PYGdefault@tok@kt\endcsname{\def\PYGdefault@tc##1{\textcolor[rgb]{0.69,0.00,0.25}{##1}}}
\expandafter\def\csname PYGdefault@tok@o\endcsname{\def\PYGdefault@tc##1{\textcolor[rgb]{0.40,0.40,0.40}{##1}}}
\expandafter\def\csname PYGdefault@tok@ow\endcsname{\let\PYGdefault@bf=\textbf\def\PYGdefault@tc##1{\textcolor[rgb]{0.67,0.13,1.00}{##1}}}
\expandafter\def\csname PYGdefault@tok@nb\endcsname{\def\PYGdefault@tc##1{\textcolor[rgb]{0.00,0.50,0.00}{##1}}}
\expandafter\def\csname PYGdefault@tok@nf\endcsname{\def\PYGdefault@tc##1{\textcolor[rgb]{0.00,0.00,1.00}{##1}}}
\expandafter\def\csname PYGdefault@tok@nc\endcsname{\let\PYGdefault@bf=\textbf\def\PYGdefault@tc##1{\textcolor[rgb]{0.00,0.00,1.00}{##1}}}
\expandafter\def\csname PYGdefault@tok@nn\endcsname{\let\PYGdefault@bf=\textbf\def\PYGdefault@tc##1{\textcolor[rgb]{0.00,0.00,1.00}{##1}}}
\expandafter\def\csname PYGdefault@tok@ne\endcsname{\let\PYGdefault@bf=\textbf\def\PYGdefault@tc##1{\textcolor[rgb]{0.82,0.25,0.23}{##1}}}
\expandafter\def\csname PYGdefault@tok@nv\endcsname{\def\PYGdefault@tc##1{\textcolor[rgb]{0.10,0.09,0.49}{##1}}}
\expandafter\def\csname PYGdefault@tok@no\endcsname{\def\PYGdefault@tc##1{\textcolor[rgb]{0.53,0.00,0.00}{##1}}}
\expandafter\def\csname PYGdefault@tok@nl\endcsname{\def\PYGdefault@tc##1{\textcolor[rgb]{0.63,0.63,0.00}{##1}}}
\expandafter\def\csname PYGdefault@tok@ni\endcsname{\let\PYGdefault@bf=\textbf\def\PYGdefault@tc##1{\textcolor[rgb]{0.60,0.60,0.60}{##1}}}
\expandafter\def\csname PYGdefault@tok@na\endcsname{\def\PYGdefault@tc##1{\textcolor[rgb]{0.49,0.56,0.16}{##1}}}
\expandafter\def\csname PYGdefault@tok@nt\endcsname{\let\PYGdefault@bf=\textbf\def\PYGdefault@tc##1{\textcolor[rgb]{0.00,0.50,0.00}{##1}}}
\expandafter\def\csname PYGdefault@tok@nd\endcsname{\def\PYGdefault@tc##1{\textcolor[rgb]{0.67,0.13,1.00}{##1}}}
\expandafter\def\csname PYGdefault@tok@s\endcsname{\def\PYGdefault@tc##1{\textcolor[rgb]{0.73,0.13,0.13}{##1}}}
\expandafter\def\csname PYGdefault@tok@sd\endcsname{\let\PYGdefault@it=\textit\def\PYGdefault@tc##1{\textcolor[rgb]{0.73,0.13,0.13}{##1}}}
\expandafter\def\csname PYGdefault@tok@si\endcsname{\let\PYGdefault@bf=\textbf\def\PYGdefault@tc##1{\textcolor[rgb]{0.73,0.40,0.53}{##1}}}
\expandafter\def\csname PYGdefault@tok@se\endcsname{\let\PYGdefault@bf=\textbf\def\PYGdefault@tc##1{\textcolor[rgb]{0.73,0.40,0.13}{##1}}}
\expandafter\def\csname PYGdefault@tok@sr\endcsname{\def\PYGdefault@tc##1{\textcolor[rgb]{0.73,0.40,0.53}{##1}}}
\expandafter\def\csname PYGdefault@tok@ss\endcsname{\def\PYGdefault@tc##1{\textcolor[rgb]{0.10,0.09,0.49}{##1}}}
\expandafter\def\csname PYGdefault@tok@sx\endcsname{\def\PYGdefault@tc##1{\textcolor[rgb]{0.00,0.50,0.00}{##1}}}
\expandafter\def\csname PYGdefault@tok@m\endcsname{\def\PYGdefault@tc##1{\textcolor[rgb]{0.40,0.40,0.40}{##1}}}
\expandafter\def\csname PYGdefault@tok@gh\endcsname{\let\PYGdefault@bf=\textbf\def\PYGdefault@tc##1{\textcolor[rgb]{0.00,0.00,0.50}{##1}}}
\expandafter\def\csname PYGdefault@tok@gu\endcsname{\let\PYGdefault@bf=\textbf\def\PYGdefault@tc##1{\textcolor[rgb]{0.50,0.00,0.50}{##1}}}
\expandafter\def\csname PYGdefault@tok@gd\endcsname{\def\PYGdefault@tc##1{\textcolor[rgb]{0.63,0.00,0.00}{##1}}}
\expandafter\def\csname PYGdefault@tok@gi\endcsname{\def\PYGdefault@tc##1{\textcolor[rgb]{0.00,0.63,0.00}{##1}}}
\expandafter\def\csname PYGdefault@tok@gr\endcsname{\def\PYGdefault@tc##1{\textcolor[rgb]{1.00,0.00,0.00}{##1}}}
\expandafter\def\csname PYGdefault@tok@ge\endcsname{\let\PYGdefault@it=\textit}
\expandafter\def\csname PYGdefault@tok@gs\endcsname{\let\PYGdefault@bf=\textbf}
\expandafter\def\csname PYGdefault@tok@gp\endcsname{\let\PYGdefault@bf=\textbf\def\PYGdefault@tc##1{\textcolor[rgb]{0.00,0.00,0.50}{##1}}}
\expandafter\def\csname PYGdefault@tok@go\endcsname{\def\PYGdefault@tc##1{\textcolor[rgb]{0.53,0.53,0.53}{##1}}}
\expandafter\def\csname PYGdefault@tok@gt\endcsname{\def\PYGdefault@tc##1{\textcolor[rgb]{0.00,0.27,0.87}{##1}}}
\expandafter\def\csname PYGdefault@tok@err\endcsname{\def\PYGdefault@bc##1{\setlength{\fboxsep}{0pt}\fcolorbox[rgb]{1.00,0.00,0.00}{1,1,1}{\strut ##1}}}
\expandafter\def\csname PYGdefault@tok@kc\endcsname{\let\PYGdefault@bf=\textbf\def\PYGdefault@tc##1{\textcolor[rgb]{0.00,0.50,0.00}{##1}}}
\expandafter\def\csname PYGdefault@tok@kd\endcsname{\let\PYGdefault@bf=\textbf\def\PYGdefault@tc##1{\textcolor[rgb]{0.00,0.50,0.00}{##1}}}
\expandafter\def\csname PYGdefault@tok@kn\endcsname{\let\PYGdefault@bf=\textbf\def\PYGdefault@tc##1{\textcolor[rgb]{0.00,0.50,0.00}{##1}}}
\expandafter\def\csname PYGdefault@tok@kr\endcsname{\let\PYGdefault@bf=\textbf\def\PYGdefault@tc##1{\textcolor[rgb]{0.00,0.50,0.00}{##1}}}
\expandafter\def\csname PYGdefault@tok@bp\endcsname{\def\PYGdefault@tc##1{\textcolor[rgb]{0.00,0.50,0.00}{##1}}}
\expandafter\def\csname PYGdefault@tok@fm\endcsname{\def\PYGdefault@tc##1{\textcolor[rgb]{0.00,0.00,1.00}{##1}}}
\expandafter\def\csname PYGdefault@tok@vc\endcsname{\def\PYGdefault@tc##1{\textcolor[rgb]{0.10,0.09,0.49}{##1}}}
\expandafter\def\csname PYGdefault@tok@vg\endcsname{\def\PYGdefault@tc##1{\textcolor[rgb]{0.10,0.09,0.49}{##1}}}
\expandafter\def\csname PYGdefault@tok@vi\endcsname{\def\PYGdefault@tc##1{\textcolor[rgb]{0.10,0.09,0.49}{##1}}}
\expandafter\def\csname PYGdefault@tok@vm\endcsname{\def\PYGdefault@tc##1{\textcolor[rgb]{0.10,0.09,0.49}{##1}}}
\expandafter\def\csname PYGdefault@tok@sa\endcsname{\def\PYGdefault@tc##1{\textcolor[rgb]{0.73,0.13,0.13}{##1}}}
\expandafter\def\csname PYGdefault@tok@sb\endcsname{\def\PYGdefault@tc##1{\textcolor[rgb]{0.73,0.13,0.13}{##1}}}
\expandafter\def\csname PYGdefault@tok@sc\endcsname{\def\PYGdefault@tc##1{\textcolor[rgb]{0.73,0.13,0.13}{##1}}}
\expandafter\def\csname PYGdefault@tok@dl\endcsname{\def\PYGdefault@tc##1{\textcolor[rgb]{0.73,0.13,0.13}{##1}}}
\expandafter\def\csname PYGdefault@tok@s2\endcsname{\def\PYGdefault@tc##1{\textcolor[rgb]{0.73,0.13,0.13}{##1}}}
\expandafter\def\csname PYGdefault@tok@sh\endcsname{\def\PYGdefault@tc##1{\textcolor[rgb]{0.73,0.13,0.13}{##1}}}
\expandafter\def\csname PYGdefault@tok@s1\endcsname{\def\PYGdefault@tc##1{\textcolor[rgb]{0.73,0.13,0.13}{##1}}}
\expandafter\def\csname PYGdefault@tok@mb\endcsname{\def\PYGdefault@tc##1{\textcolor[rgb]{0.40,0.40,0.40}{##1}}}
\expandafter\def\csname PYGdefault@tok@mf\endcsname{\def\PYGdefault@tc##1{\textcolor[rgb]{0.40,0.40,0.40}{##1}}}
\expandafter\def\csname PYGdefault@tok@mh\endcsname{\def\PYGdefault@tc##1{\textcolor[rgb]{0.40,0.40,0.40}{##1}}}
\expandafter\def\csname PYGdefault@tok@mi\endcsname{\def\PYGdefault@tc##1{\textcolor[rgb]{0.40,0.40,0.40}{##1}}}
\expandafter\def\csname PYGdefault@tok@il\endcsname{\def\PYGdefault@tc##1{\textcolor[rgb]{0.40,0.40,0.40}{##1}}}
\expandafter\def\csname PYGdefault@tok@mo\endcsname{\def\PYGdefault@tc##1{\textcolor[rgb]{0.40,0.40,0.40}{##1}}}
\expandafter\def\csname PYGdefault@tok@ch\endcsname{\let\PYGdefault@it=\textit\def\PYGdefault@tc##1{\textcolor[rgb]{0.25,0.50,0.50}{##1}}}
\expandafter\def\csname PYGdefault@tok@cm\endcsname{\let\PYGdefault@it=\textit\def\PYGdefault@tc##1{\textcolor[rgb]{0.25,0.50,0.50}{##1}}}
\expandafter\def\csname PYGdefault@tok@cpf\endcsname{\let\PYGdefault@it=\textit\def\PYGdefault@tc##1{\textcolor[rgb]{0.25,0.50,0.50}{##1}}}
\expandafter\def\csname PYGdefault@tok@c1\endcsname{\let\PYGdefault@it=\textit\def\PYGdefault@tc##1{\textcolor[rgb]{0.25,0.50,0.50}{##1}}}
\expandafter\def\csname PYGdefault@tok@cs\endcsname{\let\PYGdefault@it=\textit\def\PYGdefault@tc##1{\textcolor[rgb]{0.25,0.50,0.50}{##1}}}

\makeatother

\makeatletter
\def\PYG@reset{\let\PYG@it=\relax \let\PYG@bf=\relax%
    \let\PYG@ul=\relax \let\PYG@tc=\relax%
    \let\PYG@bc=\relax \let\PYG@ff=\relax}
\def\PYG@tok#1{\csname PYG@tok@#1\endcsname}
\def\PYG@toks#1+{\ifx\relax#1\empty\else%
    \PYG@tok{#1}\expandafter\PYG@toks\fi}
\def\PYG@do#1{\PYG@bc{\PYG@tc{\PYG@ul{%
    \PYG@it{\PYG@bf{\PYG@ff{#1}}}}}}}
\def\PYG#1#2{\PYG@reset\PYG@toks#1+\relax+\PYG@do{#2}}

\expandafter\def\csname PYG@tok@w\endcsname{\def\PYG@tc##1{\textcolor[rgb]{0.73,0.73,0.73}{##1}}}
\expandafter\def\csname PYG@tok@c\endcsname{\let\PYG@it=\textit\def\PYG@tc##1{\textcolor[rgb]{0.25,0.50,0.50}{##1}}}
\expandafter\def\csname PYG@tok@cp\endcsname{\def\PYG@tc##1{\textcolor[rgb]{0.74,0.48,0.00}{##1}}}
\expandafter\def\csname PYG@tok@k\endcsname{\let\PYG@bf=\textbf\def\PYG@tc##1{\textcolor[rgb]{0.00,0.50,0.00}{##1}}}
\expandafter\def\csname PYG@tok@kp\endcsname{\def\PYG@tc##1{\textcolor[rgb]{0.00,0.50,0.00}{##1}}}
\expandafter\def\csname PYG@tok@kt\endcsname{\def\PYG@tc##1{\textcolor[rgb]{0.69,0.00,0.25}{##1}}}
\expandafter\def\csname PYG@tok@o\endcsname{\def\PYG@tc##1{\textcolor[rgb]{0.40,0.40,0.40}{##1}}}
\expandafter\def\csname PYG@tok@ow\endcsname{\let\PYG@bf=\textbf\def\PYG@tc##1{\textcolor[rgb]{0.67,0.13,1.00}{##1}}}
\expandafter\def\csname PYG@tok@nb\endcsname{\def\PYG@tc##1{\textcolor[rgb]{0.00,0.50,0.00}{##1}}}
\expandafter\def\csname PYG@tok@nf\endcsname{\def\PYG@tc##1{\textcolor[rgb]{0.00,0.00,1.00}{##1}}}
\expandafter\def\csname PYG@tok@nc\endcsname{\let\PYG@bf=\textbf\def\PYG@tc##1{\textcolor[rgb]{0.00,0.00,1.00}{##1}}}
\expandafter\def\csname PYG@tok@nn\endcsname{\let\PYG@bf=\textbf\def\PYG@tc##1{\textcolor[rgb]{0.00,0.00,1.00}{##1}}}
\expandafter\def\csname PYG@tok@ne\endcsname{\let\PYG@bf=\textbf\def\PYG@tc##1{\textcolor[rgb]{0.82,0.25,0.23}{##1}}}
\expandafter\def\csname PYG@tok@nv\endcsname{\def\PYG@tc##1{\textcolor[rgb]{0.10,0.09,0.49}{##1}}}
\expandafter\def\csname PYG@tok@no\endcsname{\def\PYG@tc##1{\textcolor[rgb]{0.53,0.00,0.00}{##1}}}
\expandafter\def\csname PYG@tok@nl\endcsname{\def\PYG@tc##1{\textcolor[rgb]{0.63,0.63,0.00}{##1}}}
\expandafter\def\csname PYG@tok@ni\endcsname{\let\PYG@bf=\textbf\def\PYG@tc##1{\textcolor[rgb]{0.60,0.60,0.60}{##1}}}
\expandafter\def\csname PYG@tok@na\endcsname{\def\PYG@tc##1{\textcolor[rgb]{0.49,0.56,0.16}{##1}}}
\expandafter\def\csname PYG@tok@nt\endcsname{\let\PYG@bf=\textbf\def\PYG@tc##1{\textcolor[rgb]{0.00,0.50,0.00}{##1}}}
\expandafter\def\csname PYG@tok@nd\endcsname{\def\PYG@tc##1{\textcolor[rgb]{0.67,0.13,1.00}{##1}}}
\expandafter\def\csname PYG@tok@s\endcsname{\def\PYG@tc##1{\textcolor[rgb]{0.73,0.13,0.13}{##1}}}
\expandafter\def\csname PYG@tok@sd\endcsname{\let\PYG@it=\textit\def\PYG@tc##1{\textcolor[rgb]{0.73,0.13,0.13}{##1}}}
\expandafter\def\csname PYG@tok@si\endcsname{\let\PYG@bf=\textbf\def\PYG@tc##1{\textcolor[rgb]{0.73,0.40,0.53}{##1}}}
\expandafter\def\csname PYG@tok@se\endcsname{\let\PYG@bf=\textbf\def\PYG@tc##1{\textcolor[rgb]{0.73,0.40,0.13}{##1}}}
\expandafter\def\csname PYG@tok@sr\endcsname{\def\PYG@tc##1{\textcolor[rgb]{0.73,0.40,0.53}{##1}}}
\expandafter\def\csname PYG@tok@ss\endcsname{\def\PYG@tc##1{\textcolor[rgb]{0.10,0.09,0.49}{##1}}}
\expandafter\def\csname PYG@tok@sx\endcsname{\def\PYG@tc##1{\textcolor[rgb]{0.00,0.50,0.00}{##1}}}
\expandafter\def\csname PYG@tok@m\endcsname{\def\PYG@tc##1{\textcolor[rgb]{0.40,0.40,0.40}{##1}}}
\expandafter\def\csname PYG@tok@gh\endcsname{\let\PYG@bf=\textbf\def\PYG@tc##1{\textcolor[rgb]{0.00,0.00,0.50}{##1}}}
\expandafter\def\csname PYG@tok@gu\endcsname{\let\PYG@bf=\textbf\def\PYG@tc##1{\textcolor[rgb]{0.50,0.00,0.50}{##1}}}
\expandafter\def\csname PYG@tok@gd\endcsname{\def\PYG@tc##1{\textcolor[rgb]{0.63,0.00,0.00}{##1}}}
\expandafter\def\csname PYG@tok@gi\endcsname{\def\PYG@tc##1{\textcolor[rgb]{0.00,0.63,0.00}{##1}}}
\expandafter\def\csname PYG@tok@gr\endcsname{\def\PYG@tc##1{\textcolor[rgb]{1.00,0.00,0.00}{##1}}}
\expandafter\def\csname PYG@tok@ge\endcsname{\let\PYG@it=\textit}
\expandafter\def\csname PYG@tok@gs\endcsname{\let\PYG@bf=\textbf}
\expandafter\def\csname PYG@tok@gp\endcsname{\let\PYG@bf=\textbf\def\PYG@tc##1{\textcolor[rgb]{0.00,0.00,0.50}{##1}}}
\expandafter\def\csname PYG@tok@go\endcsname{\def\PYG@tc##1{\textcolor[rgb]{0.53,0.53,0.53}{##1}}}
\expandafter\def\csname PYG@tok@gt\endcsname{\def\PYG@tc##1{\textcolor[rgb]{0.00,0.27,0.87}{##1}}}
\expandafter\def\csname PYG@tok@err\endcsname{\def\PYG@bc##1{\setlength{\fboxsep}{0pt}\fcolorbox[rgb]{1.00,0.00,0.00}{1,1,1}{\strut ##1}}}
\expandafter\def\csname PYG@tok@kc\endcsname{\let\PYG@bf=\textbf\def\PYG@tc##1{\textcolor[rgb]{0.00,0.50,0.00}{##1}}}
\expandafter\def\csname PYG@tok@kd\endcsname{\let\PYG@bf=\textbf\def\PYG@tc##1{\textcolor[rgb]{0.00,0.50,0.00}{##1}}}
\expandafter\def\csname PYG@tok@kn\endcsname{\let\PYG@bf=\textbf\def\PYG@tc##1{\textcolor[rgb]{0.00,0.50,0.00}{##1}}}
\expandafter\def\csname PYG@tok@kr\endcsname{\let\PYG@bf=\textbf\def\PYG@tc##1{\textcolor[rgb]{0.00,0.50,0.00}{##1}}}
\expandafter\def\csname PYG@tok@bp\endcsname{\def\PYG@tc##1{\textcolor[rgb]{0.00,0.50,0.00}{##1}}}
\expandafter\def\csname PYG@tok@fm\endcsname{\def\PYG@tc##1{\textcolor[rgb]{0.00,0.00,1.00}{##1}}}
\expandafter\def\csname PYG@tok@vc\endcsname{\def\PYG@tc##1{\textcolor[rgb]{0.10,0.09,0.49}{##1}}}
\expandafter\def\csname PYG@tok@vg\endcsname{\def\PYG@tc##1{\textcolor[rgb]{0.10,0.09,0.49}{##1}}}
\expandafter\def\csname PYG@tok@vi\endcsname{\def\PYG@tc##1{\textcolor[rgb]{0.10,0.09,0.49}{##1}}}
\expandafter\def\csname PYG@tok@vm\endcsname{\def\PYG@tc##1{\textcolor[rgb]{0.10,0.09,0.49}{##1}}}
\expandafter\def\csname PYG@tok@sa\endcsname{\def\PYG@tc##1{\textcolor[rgb]{0.73,0.13,0.13}{##1}}}
\expandafter\def\csname PYG@tok@sb\endcsname{\def\PYG@tc##1{\textcolor[rgb]{0.73,0.13,0.13}{##1}}}
\expandafter\def\csname PYG@tok@sc\endcsname{\def\PYG@tc##1{\textcolor[rgb]{0.73,0.13,0.13}{##1}}}
\expandafter\def\csname PYG@tok@dl\endcsname{\def\PYG@tc##1{\textcolor[rgb]{0.73,0.13,0.13}{##1}}}
\expandafter\def\csname PYG@tok@s2\endcsname{\def\PYG@tc##1{\textcolor[rgb]{0.73,0.13,0.13}{##1}}}
\expandafter\def\csname PYG@tok@sh\endcsname{\def\PYG@tc##1{\textcolor[rgb]{0.73,0.13,0.13}{##1}}}
\expandafter\def\csname PYG@tok@s1\endcsname{\def\PYG@tc##1{\textcolor[rgb]{0.73,0.13,0.13}{##1}}}
\expandafter\def\csname PYG@tok@mb\endcsname{\def\PYG@tc##1{\textcolor[rgb]{0.40,0.40,0.40}{##1}}}
\expandafter\def\csname PYG@tok@mf\endcsname{\def\PYG@tc##1{\textcolor[rgb]{0.40,0.40,0.40}{##1}}}
\expandafter\def\csname PYG@tok@mh\endcsname{\def\PYG@tc##1{\textcolor[rgb]{0.40,0.40,0.40}{##1}}}
\expandafter\def\csname PYG@tok@mi\endcsname{\def\PYG@tc##1{\textcolor[rgb]{0.40,0.40,0.40}{##1}}}
\expandafter\def\csname PYG@tok@il\endcsname{\def\PYG@tc##1{\textcolor[rgb]{0.40,0.40,0.40}{##1}}}
\expandafter\def\csname PYG@tok@mo\endcsname{\def\PYG@tc##1{\textcolor[rgb]{0.40,0.40,0.40}{##1}}}
\expandafter\def\csname PYG@tok@ch\endcsname{\let\PYG@it=\textit\def\PYG@tc##1{\textcolor[rgb]{0.25,0.50,0.50}{##1}}}
\expandafter\def\csname PYG@tok@cm\endcsname{\let\PYG@it=\textit\def\PYG@tc##1{\textcolor[rgb]{0.25,0.50,0.50}{##1}}}
\expandafter\def\csname PYG@tok@cpf\endcsname{\let\PYG@it=\textit\def\PYG@tc##1{\textcolor[rgb]{0.25,0.50,0.50}{##1}}}
\expandafter\def\csname PYG@tok@c1\endcsname{\let\PYG@it=\textit\def\PYG@tc##1{\textcolor[rgb]{0.25,0.50,0.50}{##1}}}
\expandafter\def\csname PYG@tok@cs\endcsname{\let\PYG@it=\textit\def\PYG@tc##1{\textcolor[rgb]{0.25,0.50,0.50}{##1}}}

\def\PYGZob{\char`\{}
\def\PYGZcb{\char`\}}

\def\PYGZgt{\char`\>}

\def\PYGZhy{\char`\-}

\makeatother

\setcopyright{rightsretained}
\acmPrice{}
\acmDOI{10.1145/3445814.3446725}
\acmYear{2021}
\copyrightyear{2021}
\acmSubmissionID{asplos21main-p402-p}
\acmISBN{978-1-4503-8317-2/21/04}
\acmConference[ASPLOS '21]{Proceedings of the 26th ACM International Conference on Architectural Support for Programming Languages and Operating Systems}{April 19--23, 2021}{Virtual, USA}
\acmBooktitle{Proceedings of the 26th ACM International Conference on Architectural Support for Programming Languages and Operating Systems (ASPLOS '21), April 19--23, 2021, Virtual, USA}

\begin{document}

\title{LifeStream: A High-Performance Stream Processing Engine for Periodic Streams}

\author{Anand Jayarajan}
\affiliation{%
  \institution{University of Toronto, Vector Institute}
  \country{Canada}
}
\email{anandj@cs.toronto.edu}

\author{Kimberly Hau}
\affiliation{%
  \institution{University of Toronto}
  \country{Canada}
}
\email{kimberly.hau@mail.utoronto.ca}

\author{Andrew Goodwin}
\affiliation{%
  \institution{SickKids Hospital, University of Sydney}
  \country{Canada, Australia}
}
\email{andrew.goodwin@sickkids.ca}

\author{Gennady Pekhimenko}
\affiliation{%
  \institution{University of Toronto, Vector Institute}
  \country{Canada}
}
\email{pekhimenko@cs.toronto.edu}


\begin{abstract}
Hospitals around the world collect massive amounts of physiological data from their patients every day. Recently, there has been an increase in research interest to subject this data to statistical analysis to gain more insights and provide improved medical diagnoses. Such analyses require complex computations on large volumes of data, demanding efficient data processing systems. This paper shows that currently available data processing solutions either fail to meet the performance requirements or lack simple and flexible programming interfaces. To address this problem, we propose \emph{\streamered}, a high-performance stream processing engine for physiological data. \streamer hits the sweet spot between ease of programming by providing a rich temporal query language support and performance by employing optimizations that exploit the periodic nature of physiological data. We demonstrate that \streamer achieves end-to-end performance up to $7.5\times$ higher than state-of-the-art streaming engines and $3.2\times$ than hand-optimized numerical libraries on real-world datasets and workloads.
\end{abstract}



\maketitle

\section{Introduction}\label{sec:intro}
In recent years, the healthcare industry has been experiencing an increasing trend in the adoption of approaches like data-driven diagnostic methods~\cite{mldiag, mldiab}, automated patient monitoring systems~\cite{automon}, and AI-assisted risk prediction models~\cite{drugresp, canpred, cardpred}. Advancements in the data collection technologies~\cite{telem} and recent developments in fields like statistics and machine learning~\cite{dl4h, ml4h} are the major enabling factors for this shift from the traditional methods used in healthcare practices. Hospitals collect and store hundreds of gigabytes of physiological data every day with the help of monitoring devices (e.g., Philips IntelliVue~\cite{phillips}) attached to the patients in the intensive care units (ICUs)~\cite{andrew, baljak2018scalable}. The monitors continuously collect physiological signals or waveforms such as arterial blood pressure (ABP), electrocardiogram (ECG), and electroencephalogram (EEG) and \emph{periodically} produce output at regular intervals in a \emph{streaming} manner. A single measurement or event in the waveform data typically contains a timestamp and a measurement value at that moment in time.

Traditionally, this data has been monitored and analyzed manually by the clinicians. However, statistical and machine learning-based algorithms can provide insights into complex data patterns that can help clinicians prepare a more precise diagnosis and personalized treatment plan~\cite{personal}. Moreover, statistical models are shown to accurately predict short and long-term trends in the physiological data such as cardiac arrest~\cite{cardpred} and sepsis risk~\cite{sepsis}. Even though data analytics on physiological data show great potential, several practical challenges need to be addressed to unleash its full potential.

Unlike other streaming datasets, raw physiological data has a high degree of \emph{noise} and \emph{discontinuities}. Therefore, the data needs to go through a series of data cleaning operations and transformations (e.g., normalization and frequency filtering) before it can be used for meaningful analyses. Additionally, in certain cases, researchers need to compute derived variables from the raw data (e.g, measuring heart rate from ECG signal or finding temporal correlation between multiple signals). Although general-purpose stream processing engines that can handle these types of computations do exist in both industry and academia (e.g., Apache Spark streaming~\cite{spark}, Apache Beam~\cite{dataflow}, and Apache Flink~\cite{flink}), we observe that they fail to be a good fit for processing physiological data for the following reason.

Most contemporary streaming engines provide simple and flexible programming interfaces with an implicit notion of event time and support for fine-grained windowing strategies that are well suited for building physiological data processing pipelines. However, most of them are designed with a distributed setup in mind and, unfortunately, exhibit poor single machine performance~\cite{scalecost}. This is generally compensated by scaling up the computation to large machine clusters that most hospitals neither have the infrastructure nor the required expertise to operate. Using third-party cloud services is also not widely preferred because of concerns surrounding patient privacy protection~\cite{hipaa}, security breaches~\cite{breach1, breach2}, and unpredictable downtime~\cite{outage}. This necessitates on-premise computations over limited hardware resources. Our experiments reveal that most of the contemporary streaming engines fail to provide good performance under such hardware resource-constrained setup (see Section~\ref{sec:mot} for more details).

Because of these limitations, data scientists usually prefer to write adhoc data processing pipelines using numerical libraries (e.g., NumPy~\cite{numpy}, SciPy~\cite{scipy} and Scikit-learn~\cite{sklearn}). Even though, numerical libraries provide a rich set of operations for scientific computing with efficient hand-tuned implementations, the lack of temporal ordering and windowing support, as well as the absence of the unified API specifications and common data abstraction limits data scientists' ability to efficiently program and maintain large data processing pipelines. Additionally, as the pipeline gets longer and more complex, the combined workflow's performance deteriorates due to expensive data movement across the functions and lack of cross-operation optimizations~\cite{weldpos}.

Our \emph{goal} in this work is to build a physiological data processing system that is both \emph{easy to program} and provides \emph{high performance} even under hardware resource constraints. To this end, we propose \emph{\streamered}, a new high performance stream processing engine for physiological waveform data with a rich temporal query language support. \streamer provides high performance with efficient hardware utilization using optimizations that exploit the \emph{periodic nature} of the physiological waveform data. We derive the following two key properties of temporal operations on a periodic stream:

\noindent\emph{\textbf{Linearity property: } The timestamp of the events produced by a temporal operator is a linear transformation of that of the input events.} This property lets \streamer map the events in an operator's output stream to its parent events in the input stream(s). Since all the temporal operations follow this property, the mapping can be extended to the entire lineage of every event produced during the query execution. We call this mechanism \emph{event lineage tracking}.\\
\noindent\emph{\textbf{Bounded memory footprint: } The memory footprint of a temporal operator is bounded by the size of its input(s).} Every temporal operator operates on a fixed interval size over its input streams at any given time. Since the streams' events are generated at a constant frequency, the total memory required to execute that operation can be calculated statically.

We use the above properties and propose the following three query compile-time and runtime optimizations:
\begin{enumerate}
    \item \textbf{Locality tracing: } \streamer prepares a query execution plan that maximizes the end-to-end cache locality of the pipeline by performing static analysis on the query using periodic streams' \emph{linearity property}.
    \item \textbf{Static memory allocation: } \streamer estimates the upper bound of the memory required for each operation in the pipeline using the \emph{bounded memory footprint property} and pre-allocates the memory for all intermediate results produced in the stream, thus almost eliminating the runtime memory allocation and deallocation overhead.
    \item \textbf{Targeted query processing: } We observe that the discontinuities in the physiological data are highly uneven across different signals and the number of mutually overlapping events are generally far fewer than the total number of events in the streams. Therefore, joining multiple streams together filters out many events, rendering any prior computation on them wasteful. \streamer eliminates redundant computations using \emph{event lineage tracking} mechanisms at runtime by selectively targeting only regions of input data expected to produce an output.
\end{enumerate}

We evaluate \streamer against state-of-the-art streaming engines and numerical libraries on real data analytic use cases and datasets used at The Hospital for Sick Children (SickKids)\footnote{Canada's largest pediatric hospital and the world's largest pediatric health research centre.}, Canada. On a single machine, \streamer exhibits up to $7.5\times$ higher end-to-end performance compared to the state-of-the-art streaming engine called Trill~\cite{trill}, and $3.2\times$ compared to the hand-tuned numerical libraries such as SciPy~\cite{scipy}, NumpPy~\cite{numpy}, and Scikit-learn~\cite{sklearn}. At the same time, \streamer supports a rich temporal query language that is suitable for writing diverse sets of physiological data processing pipelines based on a qualitative assessment conducted on three real-world use cases. \streamer extends the traditional query language vocabulary supported in contemporary stream processing engines to cater to certain domain-specific use cases found in the physiological data processing domain (e.g., artifact/shape detection in the signal stream). Finally, we note that even though \streamer is built for stream processing on physiological data, the ideas proposed in this paper can also be applied to other streaming use cases where data is produced at fixed intervals.\\

In summary, this paper makes the following contributions:
\begin{itemize}
    \item We showcase the challenges faced in the domain of physiological data processing and propose solutions that are evaluated on real datasets and workloads used in major hospitals.
    \item We derive two key properties of temporal operations on periodic streams, namely \emph{linearity} and \emph{bounded memory footprint}, and leverage them to propose three key optimizations, namely \emph{locality tracing}, \emph{static memory allocation}, and \emph{targeted query processing}, that can significantly improve the hardware utilization and query execution performance compared to the state-of-the-art streaming engines.
    \item We propose \streamered, a new high-performance stream processing engine with rich temporal query language support. We show that \streamer can outperform state-of-the-art streaming engines by as much as $7.5\times$ and hand-optimized numerical libraries by as much as $3.2\times$ on the end-to-end data processing pipeline on real physiological datasets.
\end{itemize}

\section{Physiological Data Collection and Processing}\label{sec:bg}
\begin{figure}
    \centering
    \includegraphics[width=\linewidth]{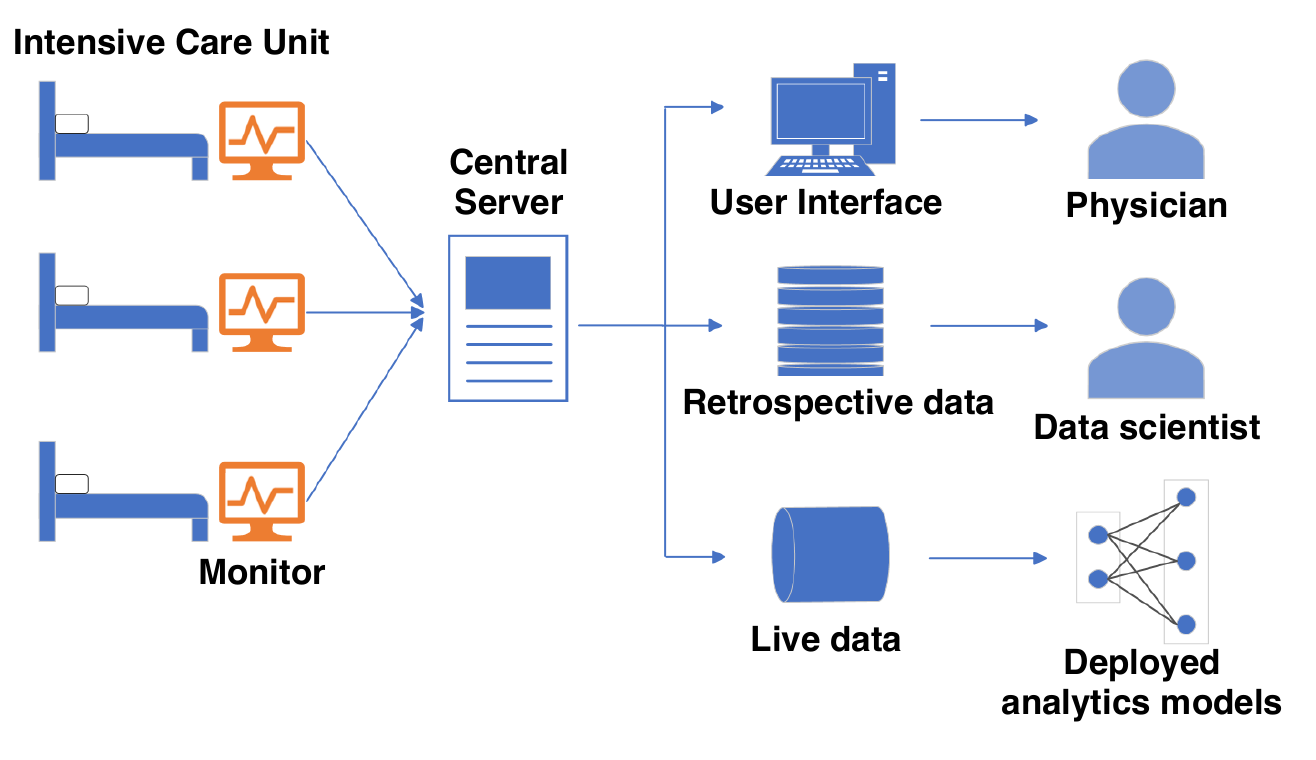}
    \caption{Typical physiological data collection infrastructure in hospitals}
    \label{fig:infra}
\end{figure}
Figure~\ref{fig:infra} shows a typical physiological data collection process from the patients in intensive care units (ICUs) to keep track of their health status with the help of multiple monitoring devices~\cite{phillips} attached to patients, each making different measurements such as arterial blood pressure (ABP), electrocardiogram (ECG), and electroencephalogram (EEG). These devices generate continuous streams of signal events at constant intervals, typically at a rate ranging from $10^{-4}$ Hz to $10^3$ Hz. Each signal event constitutes a timestamp corresponding to the time of measurement and a signal value, which is the magnitude of the measurement at that point in time. We use the term \emph{period} to refer to the shortest time interval between consecutive events in a signal. For example, a $500$ Hz signal stream would have a period of $2$ ms.

\begin{figure}
    \centering
    \includegraphics[width=\linewidth]{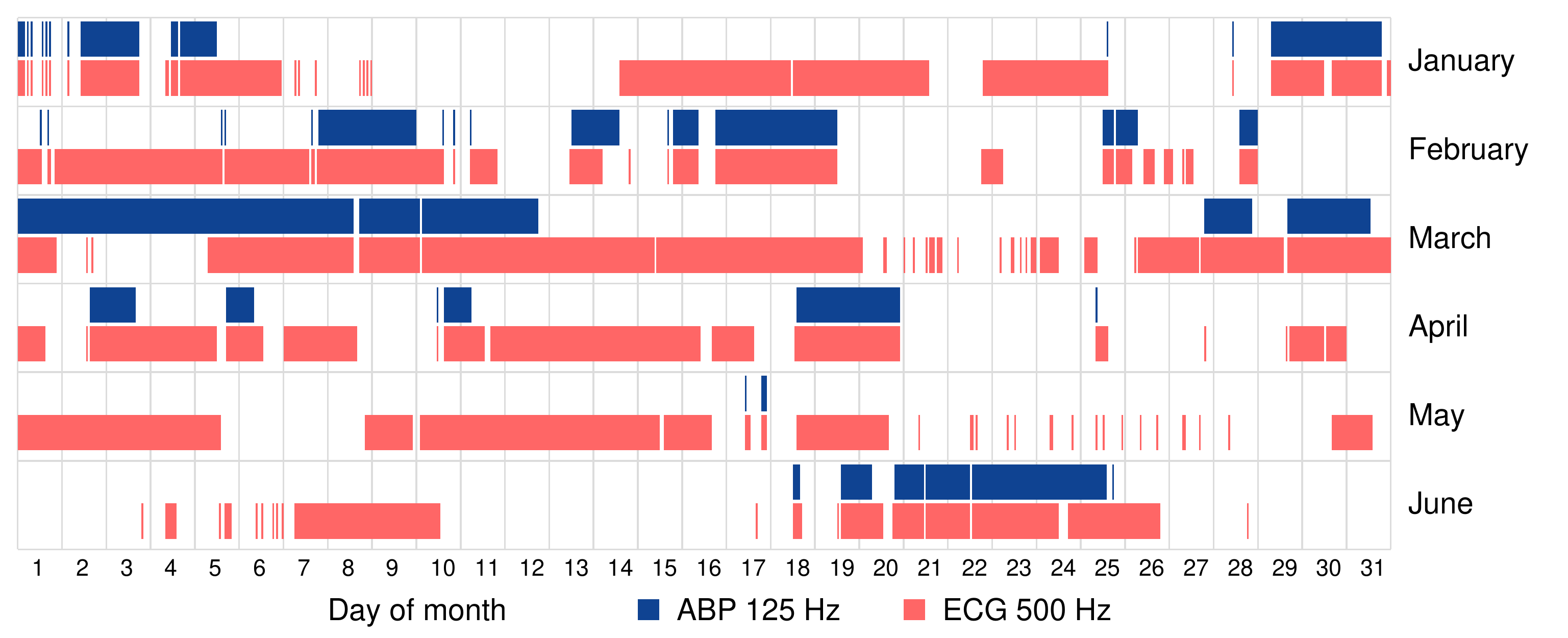}
    \caption{Distribution of ECG and ABP signals collected from a single monitoring device over first six months of 2019}
    \label{fig:disc}
\end{figure}

Unlike other streaming datasets, physiological waveform data is known to contain a high degree of \emph{noise} and many \emph{discontinuities} because of external disruptions in the connection between the monitoring devices and the patients. Figure~\ref{fig:disc} shows the discontinuities in the ECG and ABP signals collected from a single monitoring device over $6$ months. Such disruptions are common, making it virtually impossible to run meaningful analyses on top of raw data. Therefore, the data has to go through a series of transformations and data cleaning operations before a data scientist can run any data analysis. For example, standard signal processing operations like frequency-based filtering~\cite{fir} are used for removing the noise from the signals, signal value imputation methods~\cite{imp} are used to fill small discontinuities in the data stream with dummy values, and data normalization methods are used to convert all signals to a uniform scale. Additionally, different data analytics algorithms might require additional variables derived from the raw data (e.g., the systolic and diastolic blood pressure~\cite{systolic}, heart rate measured from ECG, and the temporal correlation of different signals). 

\begin{figure}[h]
    \centering
    \includegraphics[width=\linewidth]{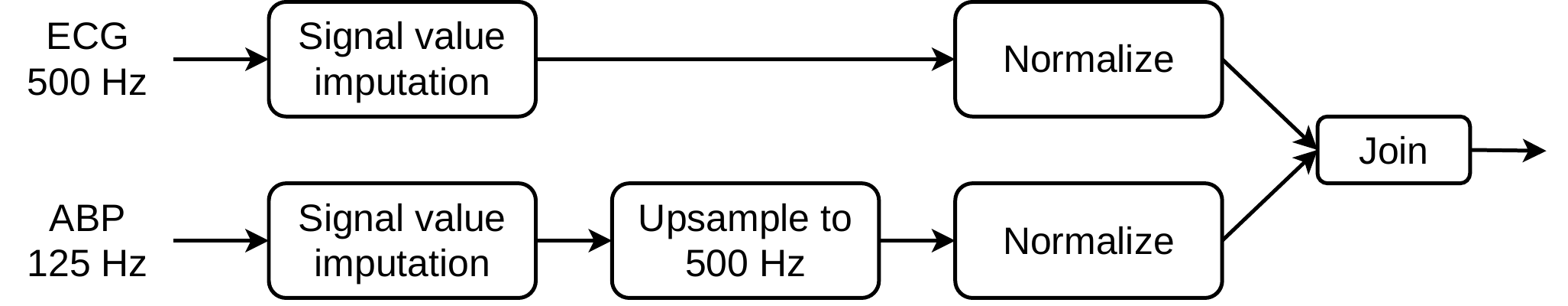}
    \caption{A sample data processing pipeline on ECG and ABP}
    \label{fig:pipeline}
\end{figure}

Figure~\ref{fig:pipeline} shows a sample data processing pipeline which joins a $125$ Hz ABP signal with $500$ Hz ECG signal based on their timestamps. First, the small gaps in both waveforms are filled using signal value imputation. Next, the ABP signal is upsampled to match the frequency of ECG. Finally, the signal values are normalized before joining them together to pair up strictly overlapping events. Even though there are several general-purpose solutions~\cite{spark,storm, flink, dataflow, trill} proposed in the past to build and process such data flow pipelines, from our experience closely working with the clinicians, data analysts, and machine learning researchers at The Hospital for Sick Children, we recognize several new challenges that make physiological waveform data processing unique.

First, the choice of operations and transformations applied on the waveform data varies considerably based on use cases. Therefore, the data scientists should have the \emph{freedom to experiment} with different data processing pipelines and should be able to do so with \emph{minimal effort}. Moreover, most of the operations and transformations applied to the waveform data follow a strict notion of the data's temporal ordering. This necessitates a flexible and easy to use programming interface with \emph{in-built temporal logic support}. Second, the data analysts and ML researchers usually first perform the experiments and analysis on the retrospective (i.e. historical) data stored in the persistent disks and then deploy their solutions on real-time data once the algorithms are finalized. The \emph{deployment must be seamless and error-free}.

Finally, since the physiological data collected are fully-identified, there are several legal restrictions~\cite{hipaa} in-place on moving the data outside hospital facilities. Even though there have been recent efforts to de-identify data for public research purposes (e.g., MIMIC~\cite{mimic}) or propose data analytics systems with strong security guarantees (e.g., Opaque~\cite{opaque}), most hospitals still prefer to keep their data private, citing the high risk associated with the leakage of patient information~\cite{risk}. Hence, the physiological data processing systems has to perform \emph{computations within limited hardware budget} available in hospitals and still provide high performance.

\section{Temporal Stream Processing for Physiological Data Analysis}\label{sec:mot}
Since physiological waveform data is produced in a streaming fashion, stream processing~\cite{stream} is a natural choice for building aforementioned data processing pipelines. Most of the modern streaming engines~\cite{spark, structstream, storm, dataflow, trill} support some temporal query language with an implicit notion of event time, temporal ordering, and fine-grained windowing strategies. To illustrate, Listing~\ref{lst:query} shows the query for a simplified version of the pipeline in Figure~\ref{fig:pipeline} which joins a $500$ Hz signal (\texttt{sig500}) and $200$ Hz signal (\texttt{sig200}) after a series of transformations.\footnote{We use signal frequencies $500$ Hz and $200$ Hz to show that \streamer can handle misaligned signals as well.} First, the signal values of \texttt{sig500} is adjusted by taking the mean of signal values on $100$ ms \emph{TumblingWindow} (fixed-size, non-overlapping, and contiguous intervals) and subtracting that from the original values. These transformed signal values are joined with the signal values from \texttt{sig200} using temporal \emph{Inner Join}. Modern streaming engines provide a simple and flexible programming interface for writing complex data processing pipelines through such query languages. However, most contemporary streaming engines are built with distributed setup in mind and, unfortunately, exhibit sub-optimal single machine performance due to low hardware utilization~\cite{scalecost, trill}. At the same time, streaming engines that are optimized for single machine performance (e.g., Trill~\cite{trill}) demonstrate orders of magnitude higher performance.

\begin{listing}
\begin{Verbatim}[commandchars=\\\{\},fontsize=\small]
\PYG{k+kt}{var} \PYG{n}{left} \PYG{p}{=} \PYG{n}{sig500}
    \PYG{p}{.}\PYG{n}{Multicast}\PYG{p}{(}\PYG{n}{s} \PYG{p}{=\PYGZgt{}} \PYG{n}{s}
      \PYG{p}{.}\PYG{n}{Select}\PYG{p}{(}\PYG{n}{e} \PYG{p}{=\PYGZgt{}} \PYG{n}{e}\PYG{p}{.}\PYG{n}{val}\PYG{p}{)} \PYG{c+c1}{// select signal value}
      \PYG{c+c1}{// compute mean and subtract from values}
      \PYG{p}{.}\PYG{n}{Join}\PYG{p}{(}\PYG{n}{s}\PYG{p}{.}\PYG{n}{TumblingWindow}\PYG{p}{(}\PYG{l+m}{100}\PYG{p}{).}\PYG{n}{Mean}\PYG{p}{(),}
        \PYG{p}{(}\PYG{n}{val}\PYG{p}{,} \PYG{n}{mean}\PYG{p}{)} \PYG{p}{=\PYGZgt{}} \PYG{n}{val} \PYG{p}{\PYGZhy{}} \PYG{n}{mean}\PYG{p}{));}
\PYG{k+kt}{var} \PYG{n}{right} \PYG{p}{=} \PYG{n}{sig200}
    \PYG{p}{.}\PYG{n}{Select}\PYG{p}{(}\PYG{n}{e} \PYG{p}{=\PYGZgt{}} \PYG{n}{e}\PYG{p}{.}\PYG{n}{val}\PYG{p}{);} \PYG{c+c1}{// select signal value}
\PYG{k+kt}{var} \PYG{n}{output} \PYG{p}{=} \PYG{n}{left}
    \PYG{c+c1}{// join with sig200 values}
    \PYG{p}{.}\PYG{n}{Join}\PYG{p}{(}\PYG{n}{right}\PYG{p}{,} \PYG{p}{(}\PYG{n}{l}\PYG{p}{,} \PYG{n}{r}\PYG{p}{)} \PYG{p}{=\PYGZgt{}} \PYG{k}{new} \PYG{p}{\PYGZob{}}\PYG{n}{l}\PYG{p}{,} \PYG{n}{r}\PYG{p}{\PYGZcb{});}
\end{Verbatim}
    \vspace{-5pt}
    \caption{Running example of a temporal query}
    \label{lst:query}
\end{listing}

To validate this in the context of waveform data processing, we compare the single-core performance of the temporal \emph{Join} operation in four major state-of-the-art streaming engines: (i) Spark streaming~\cite{spark, structstream}, (ii) Flink~\cite{flink}, (iii) Storm~\cite{storm}, and (iv) Trill~\cite{trill}. \emph{Join} is one of the most commonly used primitive operations in physiological data processing. Finding the temporal correlation of different signal streams or computing derived variables such as aggregates and joining them back with the input stream events are frequently performed computations. Therefore, the performance of \emph{Join} operation can considerably affect the entire pipeline's performance in such scenarios.

Table~\ref{tab:hw_util} shows the number of signal events joined per second by different streaming engines. We observe that Trill outperforms other streaming engines by more than $10\times$. Trill's performance benefits come from a better memory management system, improved cache locality using columnar data representation, and the use of hand-optimized primitive operators.
\begin{table}[h]
    \centering
    \caption{Throughput of Spark, Storm, Flink, Trill, and SciPy (in million events/sec)}
    \begin{tabular}{c | c c c c c}
        \hline
        \textbf{Benchmarks} & Spark & Storm & Flink & Trill & SciPy \\
        \hline
        Temporal Join & 0.07 & 0.04 & 0.09 & 0.80 & - \\
        Upsampling & - & - & - & 0.69 & 15.06 \\
        \hline
    \end{tabular}
    \vspace{5pt}
    \label{tab:hw_util}
\end{table}

Unfortunately, despite all these optimizations that made Trill significantly better than its competitors, we observe that the performance of Trill is far from being competitive with the hand-tuned implementations used by the data scientists. Such implementations are usually based on numerical libraries such as SciPy~\cite{scipy}, NumPy~\cite{numpy}, and Scikit-learn~\cite{sklearn} and provide a rich ecosystem of highly efficient data processing operators. Table~\ref{tab:hw_util} shows the performance comparison of signal upsampling~\cite{lintransf} operation implemented in Trill and the corresponding implementation available in the SciPy library. We observe that Trill is about $22\times$ slower than SciPy. This makes numerical libraries seem like a better choice for building data processing pipelines from a performance perspective. Even though this is the status quo among data scientists, we argue that such an approach has significant drawbacks from a programmability and system maintainability perspective because of the following reasons.

First, the lack of implicit notion of event time and support for flexible windowing strategies make building physiological data processing pipelines using numerical libraries significantly harder and more complicated for data scientists. For instance, writing the data transformations in Listing~\ref{lst:query} would require data scientists to manually maintain the temporal ordering of the data at the application level. Moreover, making simple tweaks like modifying the pipeline to use a rolling mean would only take a single line of change from \emph{TumblingWindow} to \emph{SlidingWindow} in a temporal query language. The same change would require a complete redesign of the code base in the typical numerical library-based approaches. 

These limitations force data scientists to make one of two undesirable choices. (i) To put considerable engineering effort to implement temporal features on top of the numerical libraries, or (ii) make compromises in their experimentation and requirements to adjust to the restrictions imposed by numeric libraries. Secondly, a lack of common API specifications and data abstractions across different libraries further complicates building large data processing pipelines, as the data scientists need to make sure the correctness of the input data types of each function and additionally perform type conversions when necessary. Such an approach quickly becomes unmanageable and error-prone as the data pipeline gets larger and more complex.

Despite all these drawbacks, data scientists still choose to go with such adhoc library-based methods for building data processing pipelines instead of more systematic approaches (e.g, using stream processing engines such as Trill) due to the significant performance benefits associated with numerical libraries. As a result, we observe that data scientists spend most of their development time writing peripheral code and extra ``glue'' logic to wire different numerical libraries together~\cite{wastingtime}, instead of focusing on their primary goal---analysing data and generating insights from it.

On the other hand, the library-based approach seems desirable from the performance perspective at first. However, prior works~\cite{weldpos, weld, split} have pointed out that even though the individual functions in these libraries may achieve high performance in isolation, they usually fail to maintain those benefits in a more complex workflow with a combination of functions because of the overhead associated with intermediate data conversion and lack of cross function optimizations. This makes numerical libraries a poor choice for building physiological data processing pipelines even from a performance standpoint.

Based on the above observations, we conclude that a physiological data processing system must provide a programming interface similar to the ones supported by the major streaming engines with flexible windowing strategies and an implicit support for event time and temporal ordering of the data. Moreover, the system must efficiently utilize the available hardware resources to provide high performance.

\section{\streamered: System Overview}\label{sec:oview}
\begin{figure}
    \centering
    \includegraphics[width=\linewidth]{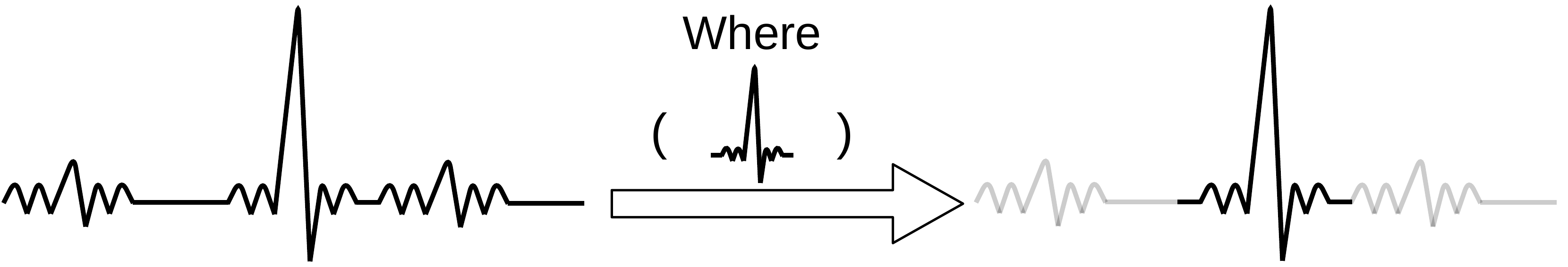}
    \caption{Shape detection using extended \emph{Where} query}
    \label{fig:shape_det}
\end{figure}
To address the challenges described in the previous section, we introduce \textbf{\streamered}, a temporal query processing engine specifically optimized for physiological waveform data processing. Compared to alternative approaches~\cite{trill, spark, scipy, sklearn, numpy}, \streamer provides: (i) superior performance by taking advantage of the periodic nature of the waveform data and optimizing the end-to-end pipeline, (ii) a rich temporal query language support~\footnote{Similar to the one provided by Trill.} with simple and flexible primitive temporal operations and fine-grained windowing support, and (iii) several extensions on traditional temporal operations that are useful for physiological data processing such as extending \emph{Where} operator to query visual patterns and shapes in data streams as shown in Figure~\ref{fig:shape_det} (see Section~\ref{sec:ap_ext} for more details).

\streamer provides the data abstraction that consists of a stream of events in chronological order. An \emph{event} is a single unit of data with three fields: (i) a user-defined \emph{payload}, (ii) a \emph{sync time}, which dictates the time from which that particular event is active, and (iii) a \emph{duration}, which  defines the active lifetime of the event. \streamer is exclusively targeting data streams in which events appear at \emph{constant intervals}. That means the sync time of every event is always at the period boundaries. Since each event's position in a stream is predictable, we use the symbolic representation of (\emph{offset}, \emph{period}) to describe a stream, where \emph{offset} is the sync time of the first event in the stream and \emph{period} is the reciprocal of the frequency. Even though we are primarily targeting physiological waveform data, any streaming data that can be represented in the aforementioned format can take advantage of the benefits of \streamered. This includes periodic streaming datasets such as performance counters produced in data centers~\cite{pingmesh}, data collected from wearable devices~\cite{wearable}, and real-time sensor data~\cite{rtsense}.

One of the key design choices we make in \streamer is to decouple the operator implementation from the data representation used for storing stream data. Most of the streaming engines, such as Trill~\cite{trill} and Spark streaming~\cite{structstream} implement their operators to take an \emph{arbitrary} sequence of events from the input stream data in the form of a batch, and produce another batch as its output. In our work, we realize that this approach has several severe limitations. First, the locality of the computation is highly tied with the batch size. Therefore, usually, the system has to trade-off the benefits of large batch processing in favor of preserving locality. Second, we observe that such operator implementations limit different compile-time and runtime optimizations we can perform on \streamer (see Section \ref{sec:opt} for more details).

To avoid such limitations, we introduce a new key construct called \emph{fixed interval sliding window} or \emph{FWindow}. FWindow is an arbitrary interval within a stream. Similar to an event, FWindow also has a sync time corresponding to the starting timestamp of the interval and a fixed size or duration. Since FWindow is an interval on a stream, the size of the FWindow should always be a multiple of the period of the stream. In \streamered, we implement all the operations based on FWindows---the operators typically take one or two FWindows as input and produce a single FWindow as output. Operators can slide the FWindows to read different parts of the stream at runtime by updating its sync time. The only restriction is that FWindows can only move forward in time to ensure monotonic progress in query execution.

\begin{table*}
    \centering
    \caption{Primitive temporal operations supported by \streamer}
    \begin{tabular}{p{0.15\linewidth} p{0.25\linewidth} p{0.1\linewidth} p{0.42\linewidth}}
        \hline
        \textbf{Operation} & \textbf{Dimension} & \textbf{Is stateful?} & \textbf{Description} \\
        \hline \hline
        Select & $[out] \leftarrow [in]$ & No & Performs a projection on the payloads of the stream events.\\
        \hline
        Where & $[out] \leftarrow [in]$ & No & Filter out events based on a predicate.\\
        \hline
        Aggregate(w, p) & $[out] \leftarrow p$ & No, if $w = p$ & Applies a user-defined aggregate function to $w$-sized windows with stride length of $p$. E.g., Sum, Max, and Mean.\\
        \hline
        Join & $[out] \leftarrow LCM([left], [right])$ & No* (Sec~\ref{sec:stateful}) & Performs temporal equijoin between two streams. E.g., InnerJoin, LeftJoin, and OuterJoin.\\
        \hline
        ClipJoin & $[out] \leftarrow [in]$ & Yes & Joins each event in one stream with immediately succeeding event in another.\\
        \hline
        Chop & $[out] \leftarrow [in]$ & Yes & Splits the intervals of each event on user-defined period boundaries.\\
        \hline
        Shift(k) & $[out] \leftarrow [in]$ & Yes & Shifts the sync time of each event by a constant.\\
        \hline
        AlterPeriod(p) & $[out] \leftarrow [in]$ & No & Alters the period of a stream.\\
        \hline
        AlterDuration & $[out] \leftarrow [in]$ & No & Alters the duration of each event in the stream.\\
        \hline
        Multicast & $[out] \leftarrow [in]$ & No & Forks a stream to allow multiply subqueries to be performed on same input stream.\\
        \hline
        Transform(w) & $[out] \leftarrow w$ & No & Applies a user-defined transformation function on $w$-sized windows that produces another $w$-sized window as output.\\
        \hline
    \end{tabular}
    \vspace{5pt}
    \label{tab:ops}
\end{table*}

Table~\ref{tab:ops} describes the set of primitive operations supported in \streamer with which data scientists can write queries on streaming data. The query is compiled into a computation graph composed of FWindows and temporal operators. The size of the FWindows are initially set to the same value as the corresponding stream's period. Figure~\ref{fig:lintrack1} shows the initial computation graph prepared from the example query in Listing~\ref{lst:query}. The FWindows are represented using a symbolic representation \emph{(offset, period)[dimension]} where the dimension is the FWindow size. This graph is then passed on through a graph transformation process to generate the final executable computation graph. Finally, the input data is streamed through the executable computation graph to generate the result.

\section{\streamered: Key Optimizations}\label{sec:opt}
\streamer maximizes resource utilization by (i) improving cache locality, (ii) reducing runtime overhead, and (iii) pruning redundant computations. We achieve this goal by identifying two key properties of the temporal operations on periodic streams described in Section~\ref{sec:prop}. Using these properties, we propose three major query compilation and execution time optimizations described in Sections~\ref{sec:loctrac} and \ref{sec:target}.

\subsection{Properties of Temporal Operations on Periodic Streams}\label{sec:prop}
\begin{figure}
    \centering
    \subfigure[Select]{
        \label{fig:select}
        \includegraphics[width=0.45\linewidth]{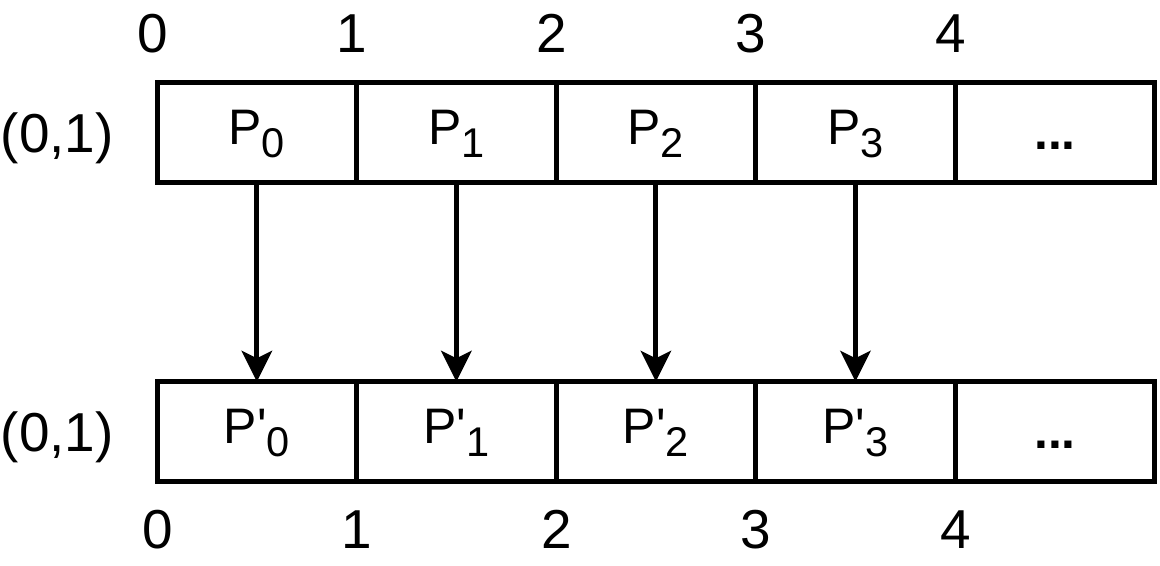}
    }
    \subfigure[Shift(k)]{
        \label{fig:shift}
        \includegraphics[width=0.465\linewidth]{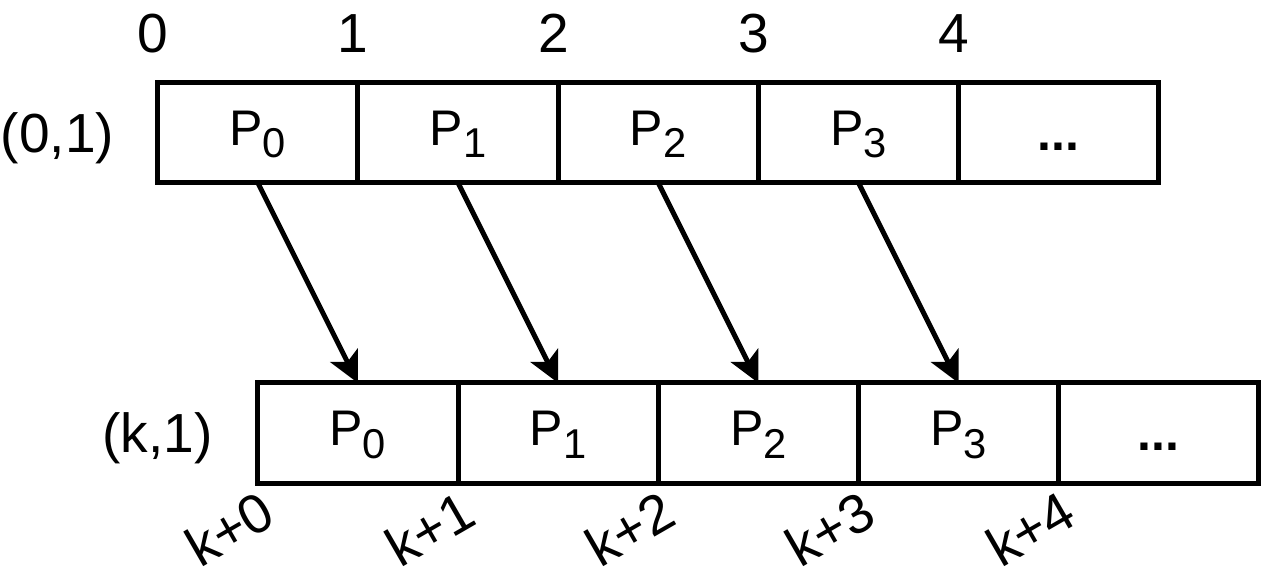}
    }%
    \\
    \subfigure[Join]{
        \label{fig:join}
        \includegraphics[width=\linewidth]{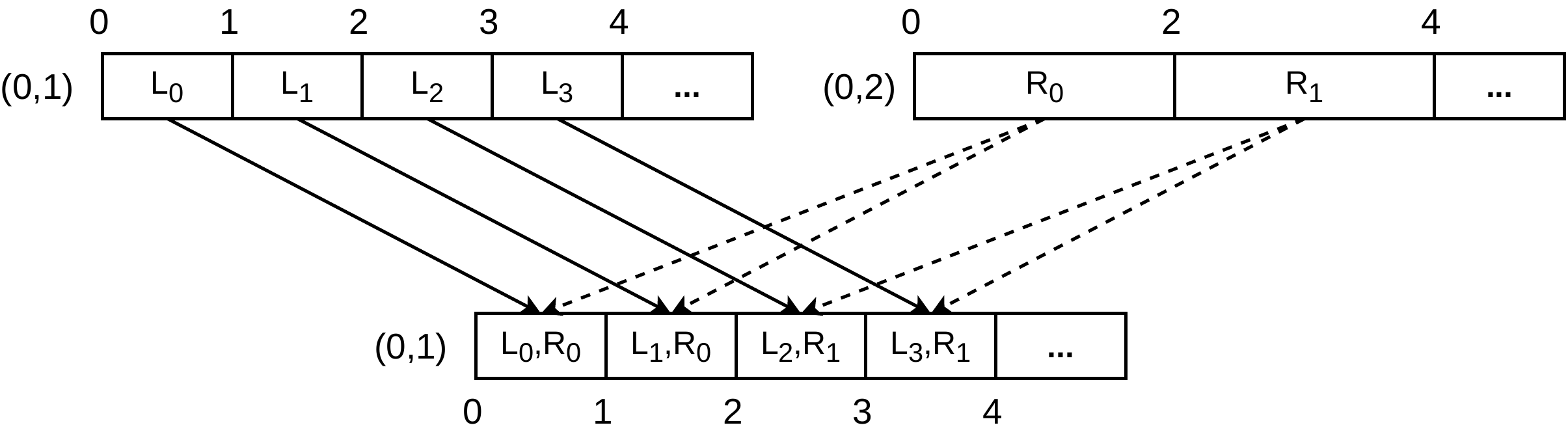}
    }
    \vspace{-10pt}
    \caption{Event lineage tracking}
    \label{fig:lintrack}
\end{figure}

\noindent\emph{\textbf{Linearity of temporal operations:} The sync time of events in the output stream of a temporal operator is a linear transformation of that of the input events.}

This property allows \streamer to map every output event of a temporal operator to the corresponding parent input event(s). Figure~\ref{fig:lintrack} shows how the event times change in the output stream of several common temporal operators such as \emph{Select}, \emph{Shift}, and \emph{Join} when applied on a periodic stream. One consequence of this property is that the period and the offset of the output stream is also a linear transformation of that of the input stream and can be computed statically. Moreover, this allows to map the events in the output stream of an operator to the corresponding parent events in the input stream(s). Figure~\ref{fig:lintrack} shows how the output events are mapped to the input events after an operator transformation. Since all temporal operators in \streamer follow this property, the mapping can be extended from the final output stream events of the query all the way to the initial input stream events. We call this mechanism \emph{event lineage tracking} and use in \streamer to improve both the cache locality and prune redundant computations.

\noindent\emph{\textbf{Bounded space complexity:} For a stream with period $p$, the maximum number of events that can be present within a given time interval $d$ is bounded by $O(d/p)$.}

One of the key properties of a periodic stream is that two events in a stream can \emph{not} overlap with each other, which means there can only be at most one event active at any point in time within a stream. Therefore, the maximum number of events in an interval is bounded by the duration of that interval. Moreover, since temporal operations also follow linearity property, all the intermediate streams generated in the query should also be periodic, and thus should also satisfy bounded space complexity property. \streamer uses this observation to estimate the maximum memory footprint of all the intermediate results and preallocate them to minimize the runtime memory allocation and deallocation overhead commonly observed in other streaming engines~\cite{tersecades,trill}.

\subsection{Locality Tracing and Memory Footprint Estimation}\label{sec:loctrac}
One thing that makes stream processing attractive is that even though the data it processes is usually massive (and sometimes can even be potentially infinite), the computations performed on the data are highly local and require only to deal with a small continuous window of events within the stream. Most streaming engines take advantage of this locality property \emph{only} at an individual operation-level and do not optimize or even maintain cross-operation locality. In \streamered, we introduce a method called \emph{locality tracing} that uses periodic streams' linearity property to precisely estimate the end-to-end locality of the computations in the \emph{entire} pipeline. Locality tracing performs static analysis on top of the computation graph and adjusts the dimensions of all the FWindows to make sure that the input and output dimensions of all the operators match. Table~\ref{tab:ops} describes the dimension translation of each supported operation in \streamered.

\begin{figure*}
    \centering
    \subfigure[Initial computation graph]{
        \label{fig:lintrack1}
        \includegraphics[width=0.19\linewidth]{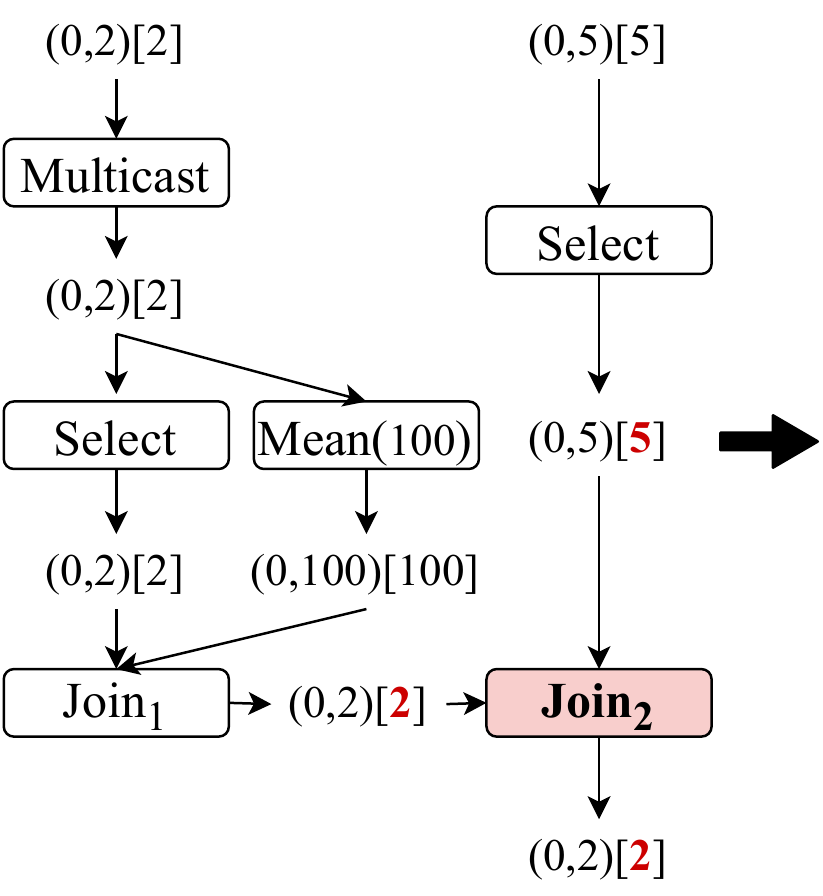}
    }
    \subfigure[Adjust $Join_2$]{
        \label{fig:lintrack2}
        \includegraphics[width=0.19\linewidth]{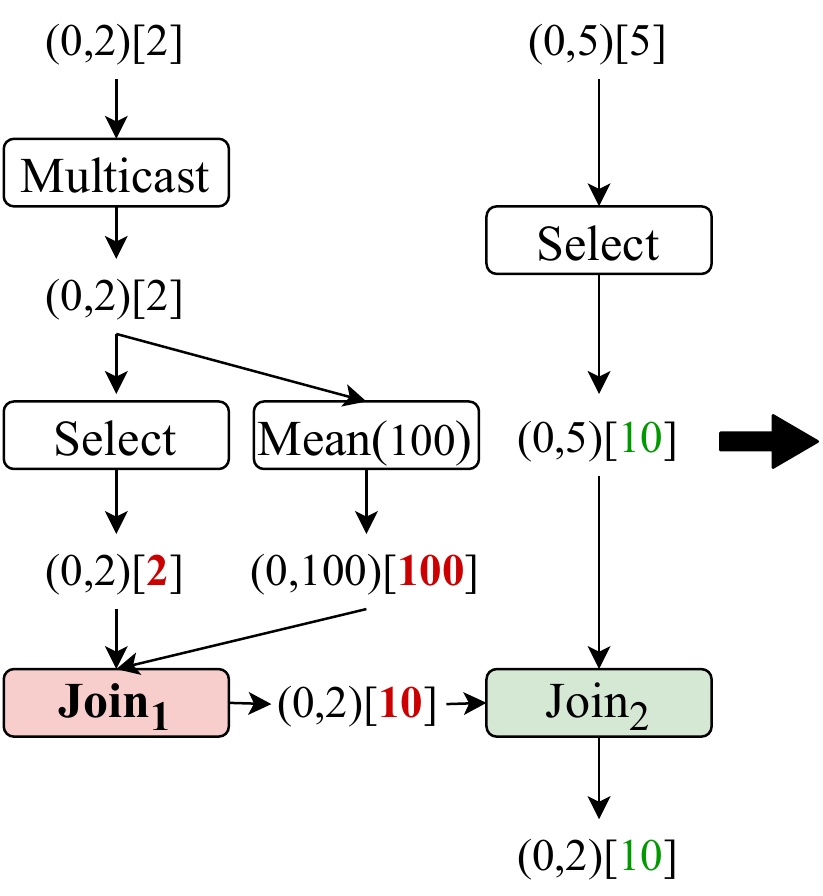}
    }
    \subfigure[Adjust $Join_1$]{
        \label{fig:lintrack3}
        \includegraphics[width=0.19\linewidth]{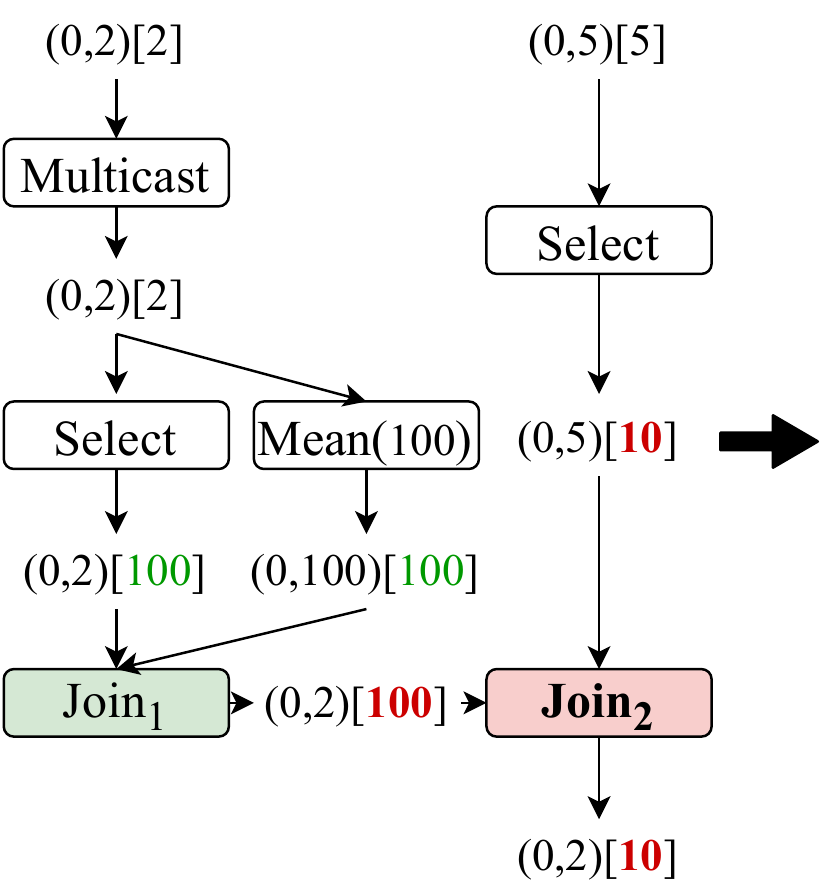}
    }
    \subfigure[Re-adjust $Join_2$]{
        \label{fig:lintrack4}
        \includegraphics[width=0.19\linewidth]{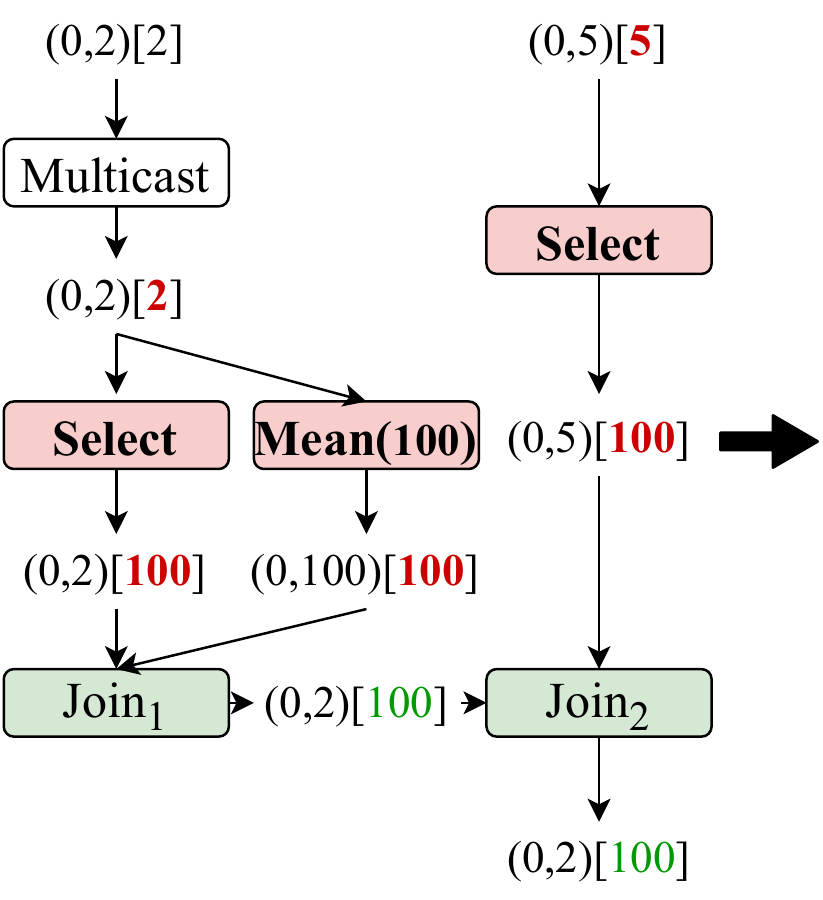}
    }
    \subfigure[Final computation graph]{
        \label{fig:lintrack6}
        \includegraphics[width=0.165\linewidth]{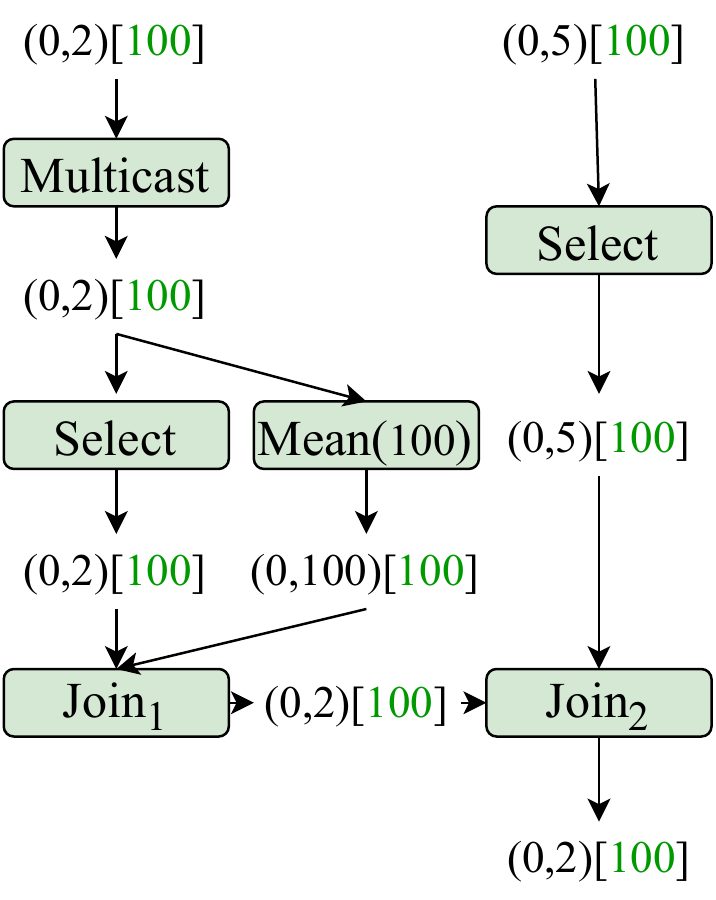}
    }
    \caption{Locality tracing procedure on the example query}
    \label{fig:locality_trace}
\end{figure*}

Figure~\ref{fig:locality_trace} shows the locality tracing procedure performed on the example query in Listing~\ref{lst:query}. The procedure starts from the end of the pipeline, and \streamer identifies a mismatch in the last \emph{Join} operation's ($Join_2$) input and output dimensions. Since the FWindow sizes has to be a constant multiple of the periods, to match the dimensions of $Join_2$, \streamer sets the FWindow sizes to the least common multiple (LCM) of the input and output dimensions. In this case, the dimensions are set to $10$. Next, the dimensions of the $Join_1$ operation are adjusted similarly. However, this adjustment introduces a mismatch in $Join_2$, which is corrected in the next step. This procedure is continued until all operations have uniform input and output dimensions. This graph transformation ensures that the intermediate results are consumed immediately by the subsequent operation(s), which, in turn, maximizes the locality of the end-to-end query.

Once the dimensions of all the operations are computed, \streamer uses the bounded space complexity property to determine the maximum memory footprint of all the FWindows. \streamer then preallocates this memory statically and keeps reusing the same memory during runtime to minimize dynamic memory allocation overhead, commonly observed in other streaming engines~\cite{tersecades}.

\subsection{Targeted Query Processing}\label{sec:target}
Most streaming engines process queries in an eager fashion where the query computation is initiated at the data ingestion side, and each subsequent operators perform the corresponding transformations on the input as soon as it receives the data and immediately pass it on to the next operator down the pipeline, irrespective of whether the next operation would need to process that data or not. In physiological data processing, this introduces many redundant computations as the data contains a high degree of discontinuity.

One of the most common examples would be the use of \emph{Inner Joins} to match overlapping events of multiple signal streams after a series of compute intensive data transformations (Figure~\ref{fig:pipeline}). Figure~\ref{fig:disc} shows that the mutually overlapping regions in ECG and ABP signal streams are far fewer than the total number of events in the individual streams. In an eager query processing model, all the events from both streams are invariably going to be passed through the intermediate transformations, even though most of them are eventually going to get discarded by the final \emph{Join} operation.

In \streamered, we address this issue by introducing a runtime optimization called targeted query processing. At runtime, \streamer uses event lineage tracking to map output FWindows of the end-to-end pipeline to corresponding parent FWindows in the input stream(s). This lets \streamer selectively target regions of the input stream(s) by sliding the FWindows and running the computations only when an output FWindow is expected to produce. Hence, targeted query processing lets \streamer skip all those compute-heavy transformations in the presence of discontinuities in the input data stream and focus only on the relevant parts of the data.

\section{\streamered: Implementation}
We implement \streamer as a library using C\# programming language and .Net core v3.1 framework~\cite{dotnet}. Hence users of \streamer can also benefit from high-level language features such as arbitrary data types, integration with custom program logic, and a rich library ecosystem~\cite{dotnetlib}. Users can also create streams from various sources, including real time data through networks, retrospective data from files, and cached data from the main memory.

As described in Section \ref{sec:oview}, the key building block that temporal operators use to access streaming data is FWindow. In FWindow, events are indexed by their sync time. In addition to the event payload, FWindow contains three extra fields: \emph{vsync}, \emph{duration}, and \emph{bitvector}. Vsync and duration fields store the sync time and duration of the events. Bitvector is used to mark the absence of an event. Every event in the FWindow has an associated bit that can either be $0$ or $1$ to indicate the event's presence or absence. All the fields in the FWindow are stored in columnar format in order to maximize cache locality as most operators only need to read from or write to a subset of the fields.

Below, we describe the details about the extended temporal query language supported in \streamer and a few implementation challenges we faced and corresponding solutions.

\subsection{Temporal Query Language Extensions}\label{sec:ap_ext}
Apart from performance benefits, \streamer also provides several additional important features through the query language extensions. We introduce a generic \emph{Transform} primitive operation which lets users write arbitrary transformations on a fixed interval of events. This operation helps users to integrate third-party libraries into the stream processing pipeline easily.

We also extend the \emph{Where} query primitive operation to support shape-based querying. As shown in Figure~\ref{fig:shape_det}, users can input arbitrary artifacts or patterns they want to detect in the stream as a list of signal values. We use a variation of the dynamic time warping (DTW) algorithm~\cite{dtw} called constrained DTW~\cite{cdtw} and re-purpose it for streaming scenario to only compute the DTW distance in linear time.

\begin{figure}
    \centering
    \includegraphics[width=\linewidth]{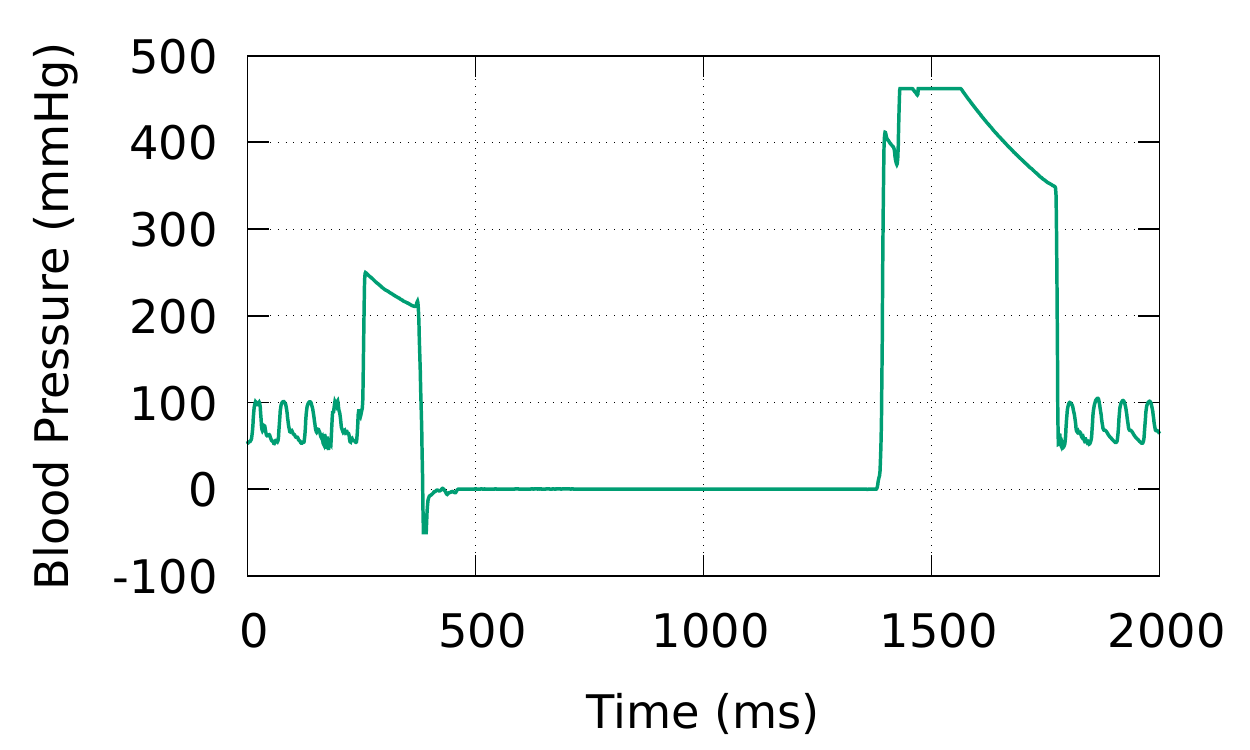}
    \caption{Line-zero artifact in arterial blood pressure (ABP)}
    \label{fig:linez}
\end{figure}
Figure~\ref{fig:linez} shows an artifact commonly found in the arterial blood pressure (ABP) signal called \emph{line-zero}, which occurs when the pressure sensors attached to patients are calibrated against atmospheric pressure. In such cases, the ABP signal values produced by the monitors would show this characteristic shape because of the distortion in the measurements. Several other artifacts are found in different signals, and data scientists generally want to remove such regions from the physiological datasets as this could negatively affect their data analysis process.

In \streamered, data scientists can use the extended \emph{Where} query primitive to filter out these artifacts from the stream by providing a representative shape as input to the query in the form of a sequence of signal values. \streamer subsequently uses the re-purposed DTW algorithm to do the pattern matching in the input stream. We measure the algorithm's accuracy over a month of ABP signal data from a single device containing $49$ line-zero artifact and achieve $0\%$ false negatives and $0.2\%$ false positives. This shows that \streamer can accurately detect such characteristic shapes in the data streams.

\subsection{FWindow Fragmentation}
Since \streamer stores and accesses data at the granularity of FWindows and they contain events of continuous fixed size interval, having small gaps in the input data might cause memory fragmentation. This might lead to low memory utilization, and in turn, low query execution performance. However, as shown in Figure~\ref{fig:disc}, most of the raw physiological data's discontinuities are generally concentrated on specific time periods rather than being randomly scattered throughout the stream. Hence most parts of the stream has continuous and densely packed sequence of data which ensures high memory utilization. However, when small gaps occur, we handle them by setting the bitvector field in the corresponding position in the FWindow to mark the absence of the events.

Another possible cause of fragmentation is when \emph{Where} query is used to filter out certain events in the stream, based on an arbitrary predicate defined by the programmer. From our experience building pipelines using \streamered, \emph{Where} operation is the least commonly used primitive operation. Even when used, it is mostly to filter out a large continuous portion of the stream (e.g., remove noisy regions or artifacts from the data). Therefore, the chances for FWindow fragmentation are minimum.

In the use cases that we evaluate in this work (see Section~\ref{sec:eval}), we observe the degree of the FWindow fragmentation is at most $0.3\%$, which is too small to impact query performance negatively.

\subsection{Stateful Temporal Operators}\label{sec:stateful}
Since the query execution in \streamer is done at the FWindow granularity and the memory allocated for the FWindows is rewritten, certain operators may need to maintain an internal state throughout the query execution (e.g., rolling aggregate operations). To handle such cases, \streamer allows operators to create constant size states during initialization, to preserve bounded memory property and ensure that there is no dynamic memory allocation during operator execution.

\begin{figure}[h]
    \centering
    \includegraphics[width=\linewidth]{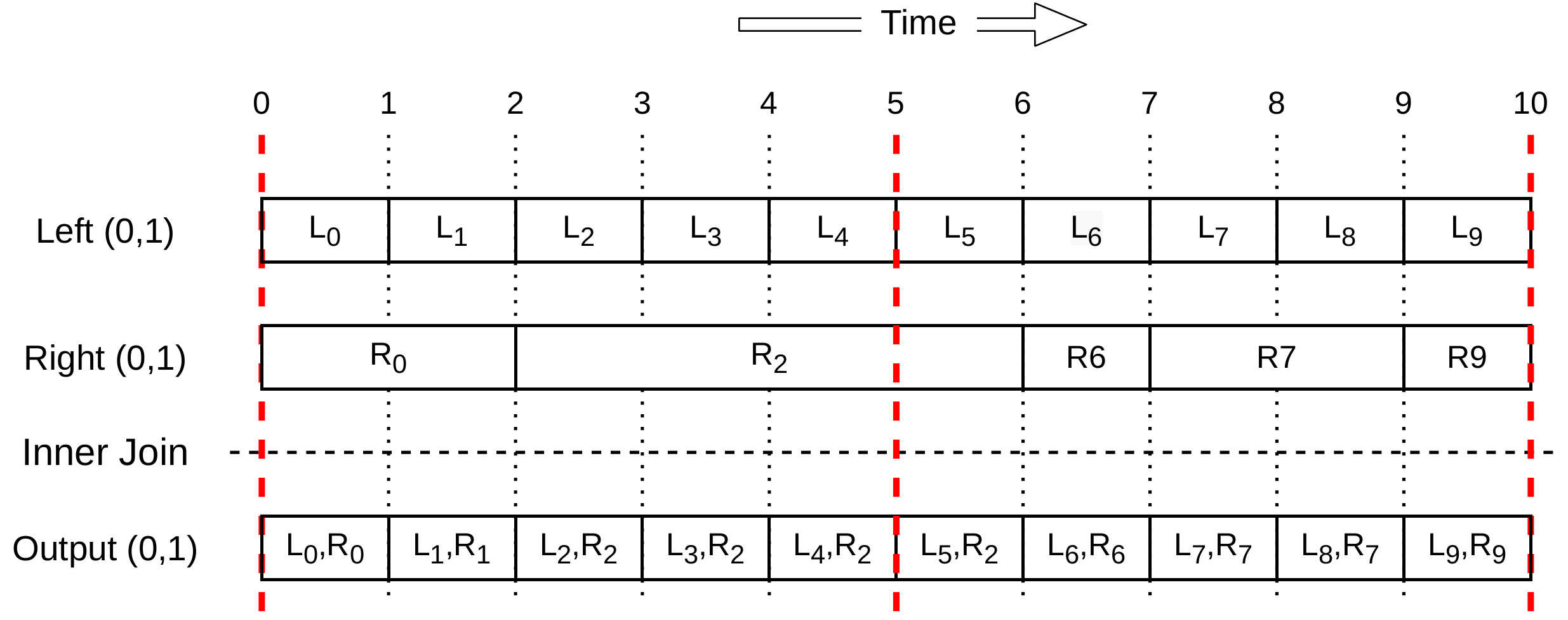}
    \caption{Stateful Join operation}
    \label{fig:state_op}
\end{figure}

Another case of a stateful operator is shown in Figure~\ref{fig:state_op}. In this case, a constant duration stream (\emph{Left}) is inner joined with another stream with arbitrary duration (\emph{Right}). The red dotted lines represent the FWindow boundaries. As shown in the figure, event $R_2$ in the \emph{Right} stream has overlapping events in the \emph{Left} stream in both FWindows. In such cases, in order for \streamer to produce the output event $(L_5,R_2)$ correctly, the \emph{Inner Join} operation needs to save the event $R_2$ in its state before moving to the second FWindow. However, the periodic nature of these streams ensures that there can only be at most one such event in a stream that can cross the interval boundary of the FWindow at any point in query execution. Therefore, the state required for temporal Join is always constant, and stateful \emph{Join} operators preserve the bounded memory property.

\section{Methodology}
\noindent\textbf{Benchmarks: } We evaluate \streamer on three categories of benchmarks. (i) \emph{Primitive micro-benchmarks}: This benchmark category includes several primitive temporal operations like \emph{Select}, \emph{Where}, \emph{Aggregate} and temporal \emph{Inner Join}. (ii) \emph{Operation benchmarks}: This includes five operations described in Table~\ref{tab:opbench} that we find commonly used by the data analysts to process physiological data. (iii) \emph{End-to-end applications}: We use the data processing pipeline described in Figure \ref{fig:pipeline} using the operations in Table~\ref{tab:opbench} as our primary benchmark for the end-to-end performance evaluation. On top of this, in Section~\ref{sec:gen}, we also use two additional pipelines used in real-world data analysis in order to show both the performance and generality of \streamered.

\begin{table}
    \centering
    \caption{Operation benchmarks and their descriptions}
    \small{
    \begin{tabular}{p{0.15\linewidth} p{0.125\linewidth} p{0.6\linewidth}}
        \hline
        \textbf{Operation} & \textbf{Libraries} & \textbf{Description} \\
        \hline \hline
        Normalize & Scikit-learn & Normalize a window of signal values using standard scores.\\
        \hline
        PassFilter & SciPy & Filter frequencies using finite impulse response~\cite{fir}.\\
        \hline
        FillConst & NumPy & Fill gaps smaller than the given window size with a constant value.\\
        \hline
        FillMean & NumPy & Fill gaps smaller than the given window size with the mean of the  values in the window.\\
        \hline
        Resample & SciPy & Up/Down sample the signal using linear interpolation~\cite{lintransf}.\\
        \hline
    \end{tabular}
    }
    \vspace{-10pt}
    \label{tab:opbench}
\end{table}

\noindent\textbf{Datasets: } Physiological waveform data we use contains signal events with a 64-bit timestamp and 32-bit floating point value. For the experiments, we use two dataset types. (i) \emph{Synthetic data}: $1000$ Hz waveform data generated for $1000$ minutes with randomly selected signal values. This dataset contains a continuous stream of signal events with no gaps. (ii) \emph{Real data}: A private dataset from The Hospital for Sick Children, containing physiological waveform data collected from $6100$ patients over the past five years. The dataset contains more than $830,000$ patient-hours of data and $250$ different signal types. However, for our experiments, we only use arterial blood pressure (ABP) and electrocardiogram (ECG) signals sampled at their default rate $125$ Hz and $500$ Hz respectively~\cite{andrew}.

\noindent\textbf{Baselines: } We compare the performance of \streamer with two baselines. (i) \emph{Microsoft Trill}, a state-of-the-art temporal query processing engine specifically optimized for single machine performance. (ii) \emph{Numeric libraries (NumLib)} like SciPy, NumPy and Scikit-learn with hand-optimized implementations for data processing operations. For end-to-end benchmarking, we implement the numerical library-based data processing pipeline in Python. To make fair performance comparisons, we tried to minimize computations done on native Python as much as possible by offloading the heavy processing to the numerical library functions. However, operations like temporal \emph{Inner Join} required pure Python implementation.

\noindent\textbf{Metrics: } We use the total execution time from a single core on a fixed input data size as the primary comparison metric for performance on primitive micro-benchmarks, operation benchmarks, and end-to-end applications. For scalability experiments, we use the throughput obtained on multiple cores/machines. Throughput is measured as the average number of signal events processed per unit time. Both execution time and throughput reported is the average of measurements from $10$ trials. The standard deviation of the measurements are observed to be less than $1\%$. For the sensitivity study on cache utilization, we use total number of last level cache (LLC) misses (median over 5 trials) on fixed workload as a comparison metric measured using Intel vTune profiler v2020~\cite{vtune}.

For all the experiments except scalability, we use 8-core (16 hyper-threaded) Intel Xeon CPU E5-2660 machine running at $2.2$ GHz, with $16$ GB RAM, and running 64-bit Ubuntu 20.04. For scalability experiments, we use up to $16$ AWS EC2 m5a.8xlarge~\cite{aws} machines each with $32$ cores and $128$ GB DRAM. We use a window size of $1$ minute for all the benchmarks unless otherwise specified.

\section{Evaluation}\label{sec:eval}
We evaluate \streamer to answer the following questions:
\begin{enumerate}
    \item How does the performance of \streamer compare to state-of-the-art streaming engines and numerical libraries?
    \item Can \streamer accelerate end-to-end performance of physiological waveform data processing pipelines?
    \item How beneficial are the proposed optimizations?
    \item How well does \streamer scale on multiple machines?
    \item Can \streamer handle diverse data analytics requirements?
\end{enumerate}

\subsection{Primitive Micro-benchmarks}
\begin{figure*}
    \centering
    \subfigure{
        \includegraphics[width=0.41\linewidth]{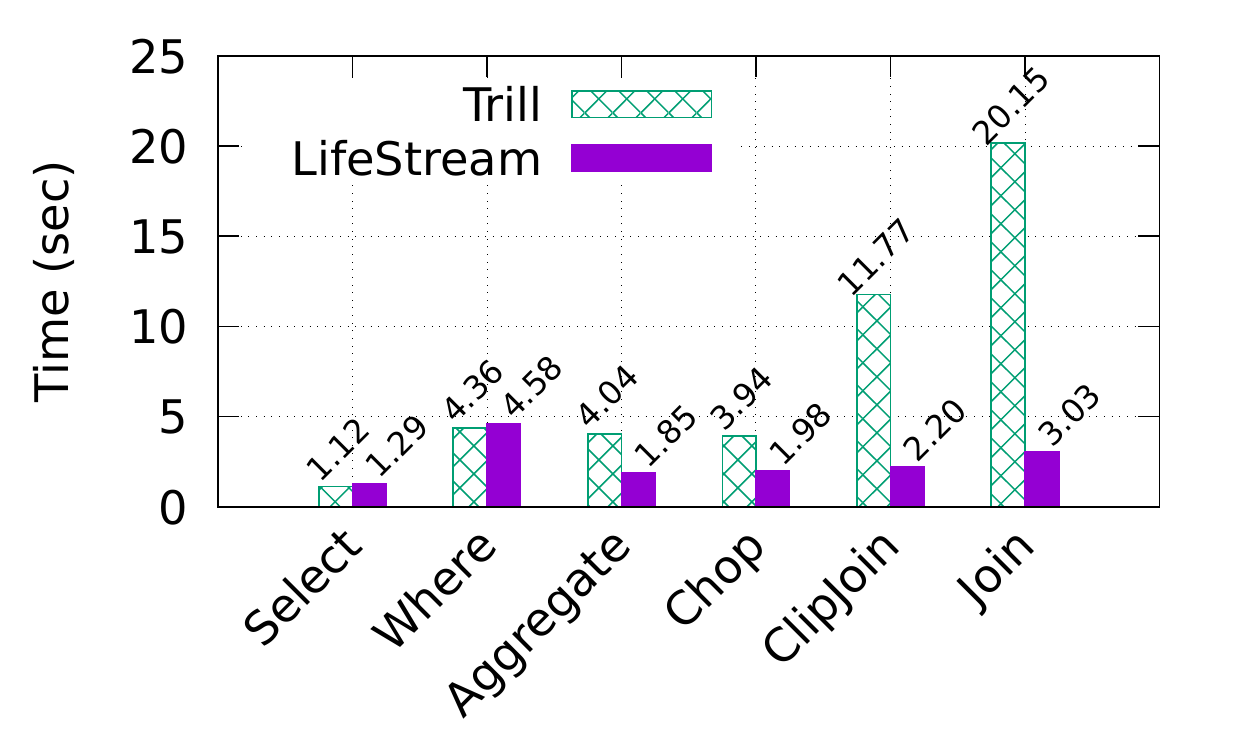}
        \label{fig:microbench}
    }
    \subfigure{
        \includegraphics[width=0.41\linewidth]{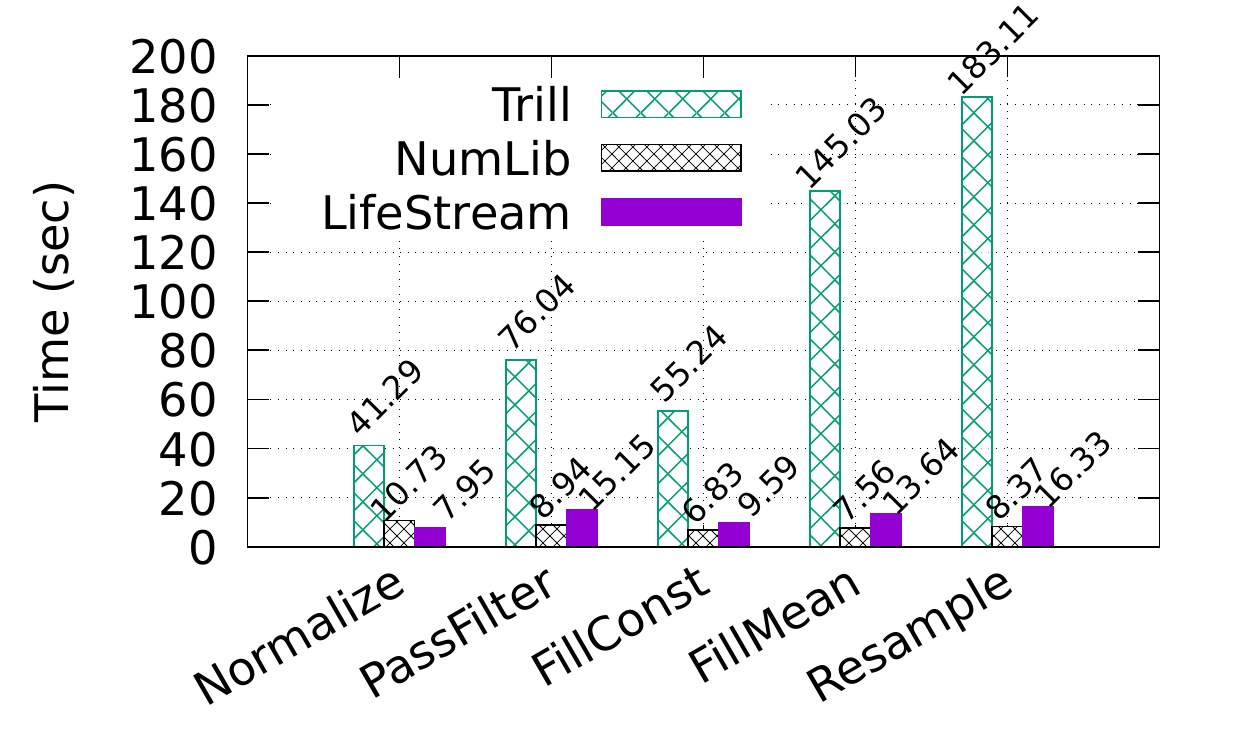}
        \label{fig:opbench}
    }
    \subfigure{
        \includegraphics[width=0.41\linewidth]{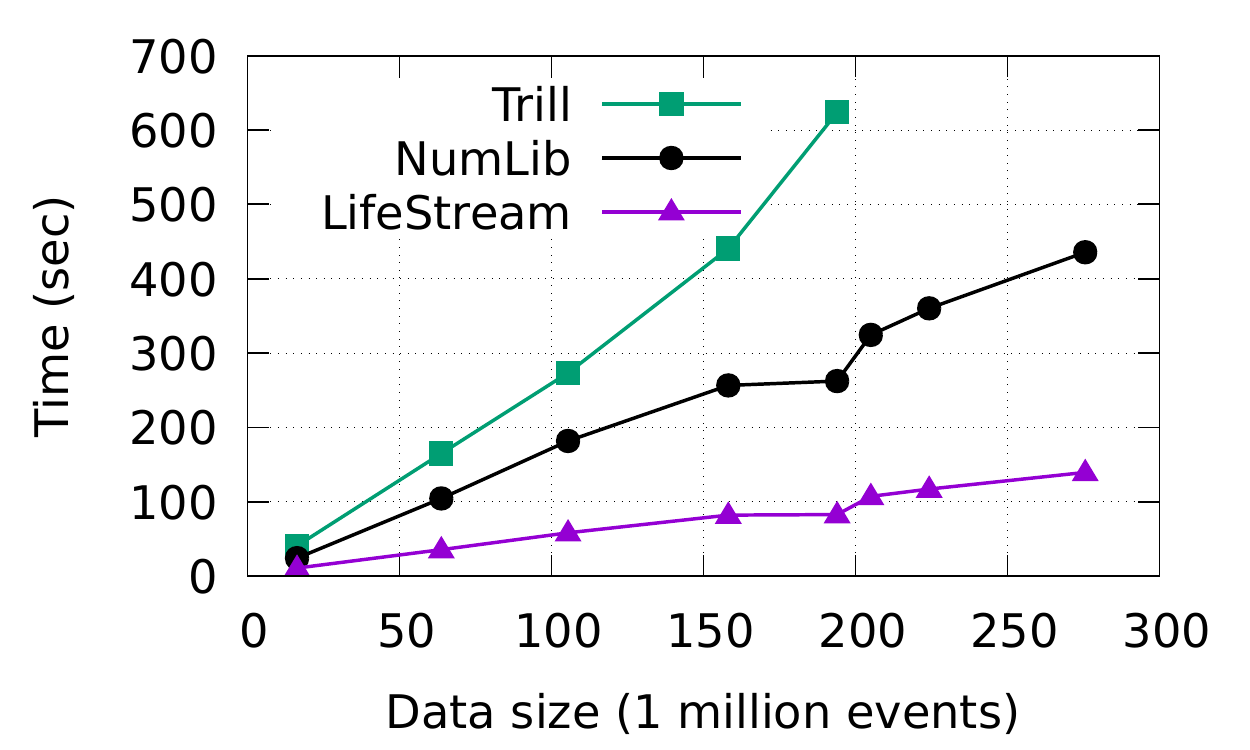}
        \label{fig:e2ebench}
    }
    \vspace{-10pt}
    \caption{(a) Primitive micro-benchmarks, (b) Operation benchmarks, (c) End-to-end applications}
\end{figure*}
We compare the performance of \streamer against \emph{Trill} on $7$ most commonly used primitive temporal operations using the synthetic dataset for these experiments. Figure~\ref{fig:microbench} shows the execution time taken by both \emph{Trill} and \emph{\streamer} and based on the results, we make the following two major conclusions.

First, on simple operations such as \emph{Select} and \emph{Where}, performance of \streamer is within $20\%$ of that of Trill. This shows that \streamer is not adding any significant overhead over already highly optimized operations in Trill. Second, we observe that \streamer shows much higher performance benefits as the operations become more complex. Operators such as  \emph{Aggregate}, \emph{Chop}, \emph{ClipJoin}, and \emph{Join} are respectively $2.17\times$, $1.98\times$, $5.34\times$, and $6.65\times$ faster than its Trill counterparts.

We attribute \streamered's high performance on primitive operations to the introduction of FWindow. FWindow greatly simplifies the operator implementations and eliminates the need for using complex data structures such as hashmaps in  temporal \emph{Join} like Trill. Moreover, since the events' timestamps and index positions are aligned, operators implemented in \streamer can compute each event's sync time from its index position without having to read from memory. Such implementation-level optimization makes \streamer efficient even at the primitive operation-level.

\subsection{Operation Benchmarks}\label{sec:opbench}
To evaluate the performance of common physiological data transformations, we implement the operations listed in Table~\ref{tab:opbench} on \streamer by writing queries using the temporal operators and compare their performance against the similar queries written in Trill and the hand-tuned implementations available in the corresponding numerical libraries specified in Table~\ref{tab:opbench}. We conduct this experiment on a $500$ Hz ECG signal from the real dataset containing $126M$ events. Figure~\ref{fig:opbench} shows the execution time of \emph{Trill}, numerical libraries (\emph{NumLib}), and \emph{\streamer} on each benchmark. We make three major conclusions from this figure.

First, across all the operations, \streamer is shown to be $5-11.21\times$ faster than Trill. This shows the effectiveness of the optimizations implemented in \streamered. Second, \streamer also exhibits comparable performance against highly optimized implementations available in the numerical libraries (within $50\%$ performance of the popular numerical libraries we evaluate). Third, in some instances such as a very commonly used \emph{Normalize} operation, \streamer even surpasses the hand-tuned performance provided by Scikit-learn library by $1.35\times$. This asserts our claim that \streamer provides ease of programming of a temporal query language without sacrificing performance.

\subsection{End-to-End Applications}\label{sec:e2ebench}
To evaluate whether \streamer can improve the end-to-end performance of data processing, we build the pipeline shown in Figure~\ref{fig:pipeline} over \streamered, Trill, and in Python using numerical libraries (NumLib). The data pipeline in all three implementations process $500$ Hz ECG and $125$ Hz ABP signals from the real dataset stored in CSV format and produces a joined signal stream after running a series of transformations on the data shown in Figure~\ref{fig:pipeline}. The dataset contains two weeks of data from a single monitoring device with $275M$ signal events. Figure~\ref{fig:e2ebench} shows the end-to-end execution time by varying the dataset size for all three implementations. We make the following two major observations from this figure.

First, \streamer outperforms both Trill and NumLib by $7.5\times$ and $3.2\times$ respectively. This reinstates that, even though individual operators in the numerical libraries can exhibit high performance when executed in isolation (as we show in Section \ref{sec:mot}), it does not necessarily translate into the best end-to-end performance due to data conversion overhead and lack of end-to-end optimizations~\cite{weldpos}.

Second, in the case of Trill, as the size of the dataset increases, the execution time rises rapidly, and Trill goes out of memory at $200M$ events. Our investigation reveals that this happens because of Trill's implementation issue with the \emph{Join} operation. Trill expects both left and right streams of the Join operation to progress at a similar pace. However, if the two streams diverge considerably, the Join operator's internal memory keeps accumulating until the available memory is exhausted. Since physiological data contains a high degree of discontinuity, it is very common for such divergence to occur during query processing. \streamer uses targeted query processing optimization (described in Section~\ref{sec:target}) to skip over such non-overlapping parts of the input data.

\subsection{Generality}\label{sec:gen}

Using \streamered, we were able to bring two data analytics models developed at the Sickkids hospital into practical deployment and real-world use, namely line-zero artifact detection model (LineZero) and cardiac arrest prediction model (CAP)~\cite{cardpred}. These models demand diverse data analytic requirements. For instance, line-zero artifact detection model performs a sliding window-based normalization operation. The cardiac arrest prediction model pipeline joins 6 different signal types together after performing operations like normalization, signal upsampling, signal value imputation, and event masking on each signal stream.

\begin{table}[h]
\centering
    \centering
    \caption{Throughput in million events per second}
    \begin{tabular}{c | c c | c}
        \hline
        \textbf{Model} & Trill & \streamer & Speedup \\
        \hline
        \textbf{LineZero} & $0.027$ & $0.315$ & $11.58\times$ \\
        \textbf{CAP} & $0.174$ & $0.877$ & $5.04\times$ \\
        \hline
    \end{tabular}
    \vspace{5pt}
    \label{tab:gen}
\end{table}
Table~\ref{tab:gen} shows the single thread performance of \streamer over Trill on LineZero and CAP models. Like the end-to-end benchmark results, \streamer achieves $11.58\times$ and $5.04\times$ speedup over Trill on LineZero and CAP respectively.

\subsection{Sensitivity Studies}
In this section, we conduct experiments to analyze the effectiveness of the optimizations applied in \streamered.\\

\noindent\textbf{Cache Utilization.} To analyze how well \streamer utilizes the cache compared to Trill using optimizations such as locality tracing, we conduct an experiment to measure the last level cache (LLC) misses of both engines on one of the most commonly used operation -- \emph{Normalize}. In order to avoid the influence of data discontinuities in the measurements, we use synthetic dataset for this experiment. We use Intel vTune~\cite{vtune} to measure the cache misses during the query execution over a constant size input dataset.

\begin{table}
\centering
    \centering
    \caption{The number of last level cache misses (in millions)}
    \begin{tabular}{c | c c c}
        \hline
        \textbf{Batch size} & $10^5$ & $10^6$ & $10^7$ \\
        \hline
        \textbf{Trill} & $2.43$ & $4.11$ & $6.73$ \\
        \textbf{\streamer} & $0.79$ & $0.82$ & $0.96$ \\
        \hline
    \end{tabular}
    \vspace{5pt}
    \label{tab:localitybench}
\end{table}

Table~\ref{tab:localitybench} shows the LLC misses in both Trill and \streamer on three different batch sizes. For a batch size of $10^5$, \streamer is experiencing $3\times$ lower cache misses as that of Trill. As the batch size increases, the number of cache misses in Trill increases significantly while \streamered's miss rate stays relatively constant. As we describe in Section~\ref{sec:loctrac}, Trill's inability to preserve cross-operation locality is the primary reason for this behaviour. The consequence of this limitation becomes more pronounced on larger batch sizes. \streamered, on the other hand, preserves the end-to-end locality of the query using locality tracing \emph{irrespective} of the batch size.\\

\begin{figure*}   
    \centering
    \subfigure{
        \includegraphics[width=0.42\linewidth]{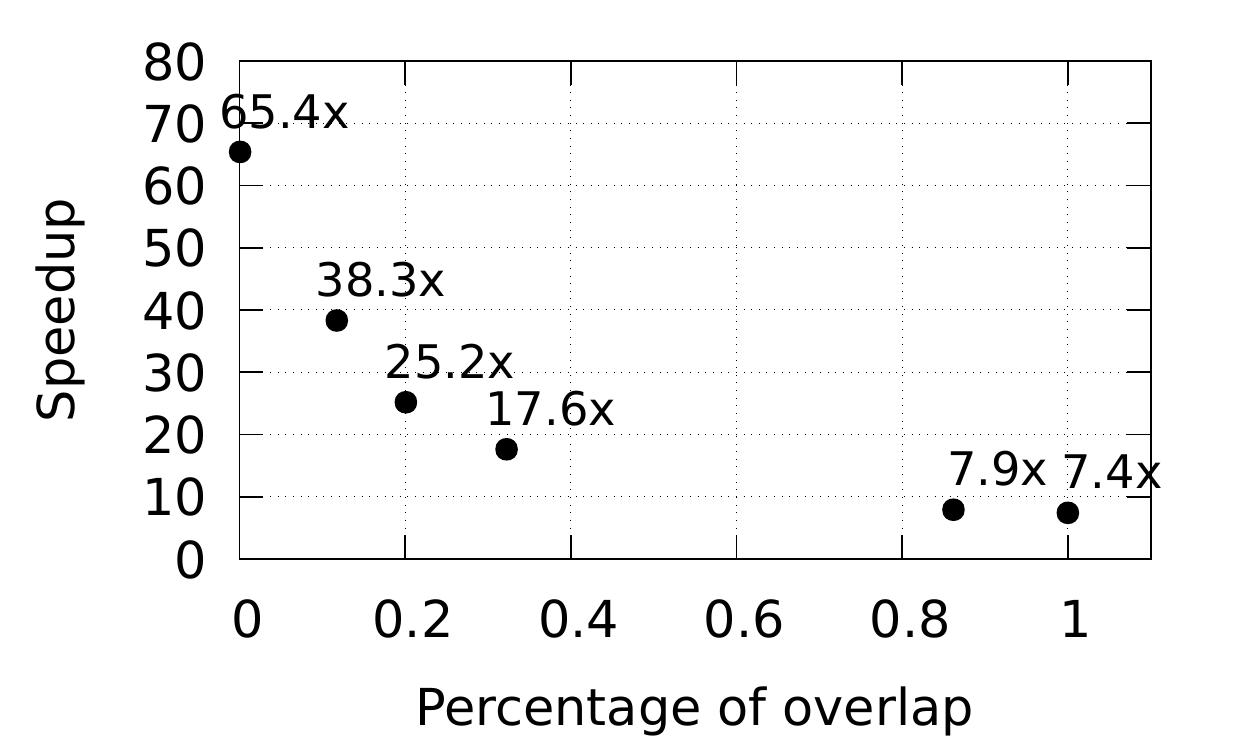}
        \label{fig:targetbench}
    }
    \subfigure{
        \includegraphics[width=0.42\linewidth]{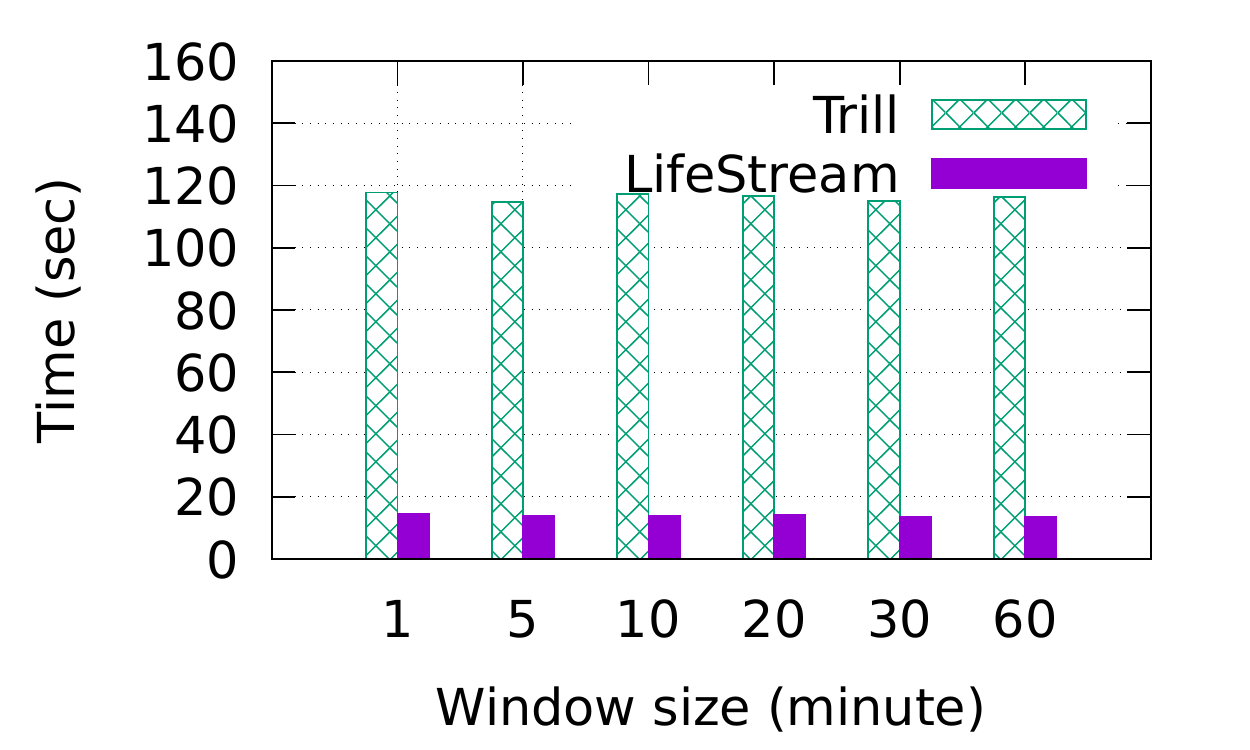}
        \label{fig:windowbench}
    }%
    \\
    \subfigure{
        \includegraphics[width=0.42\linewidth]{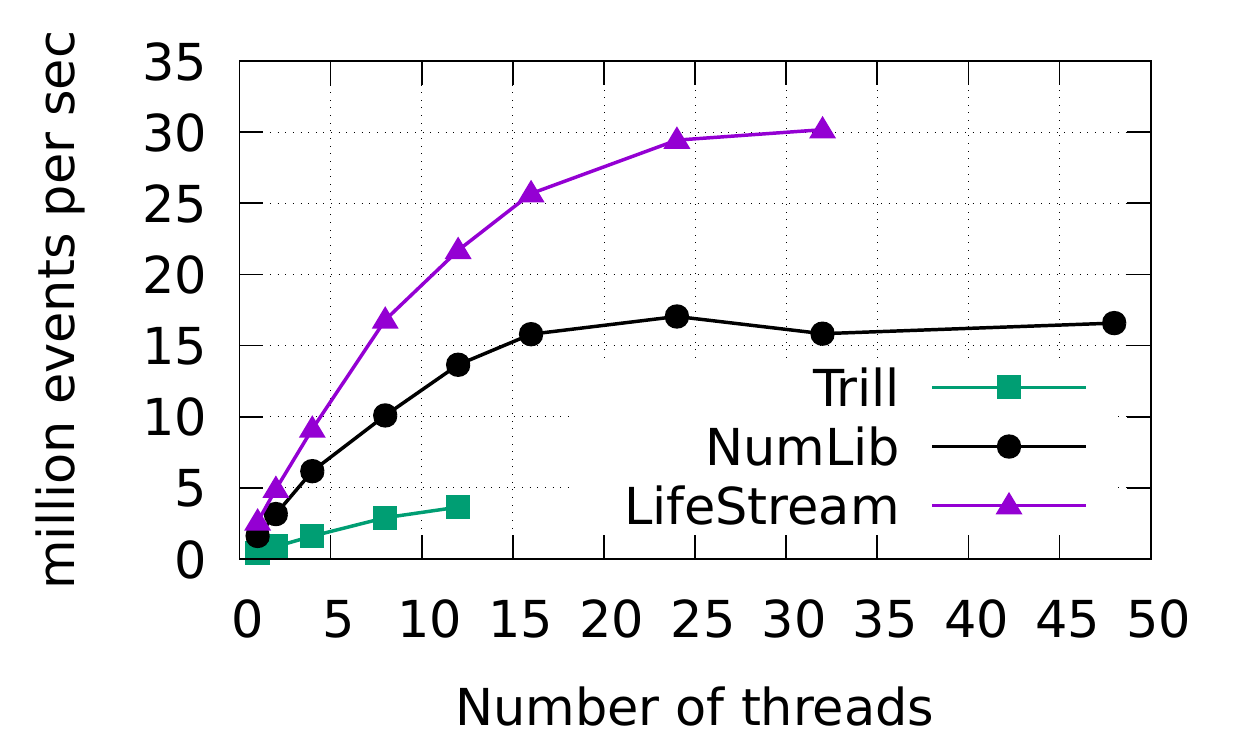}
        \label{fig:multicorebench}
    }
    \subfigure{
        \includegraphics[width=0.42\linewidth]{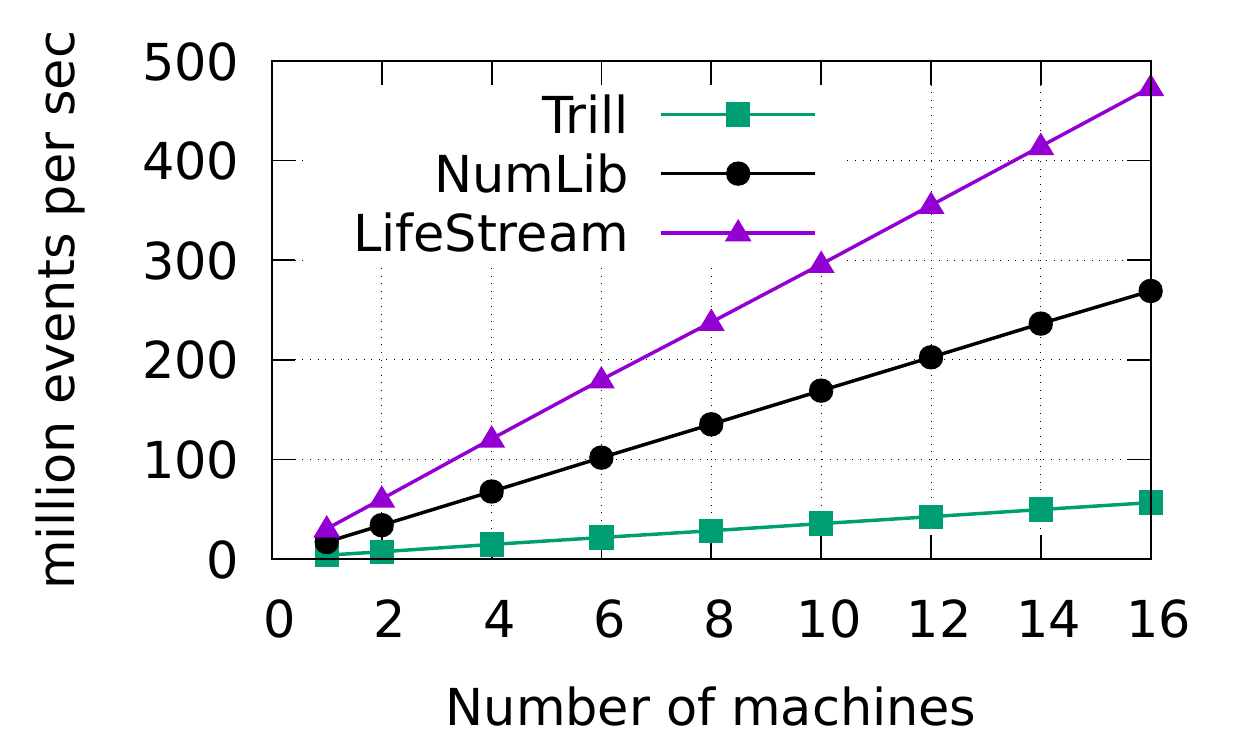}
        \label{fig:multimichbench}
    }
    \vspace{-5pt}
    \caption{(a) Targeted query processing, (b) Varying window size, (c) Multi-core scaling, (d) Multi-machine scaling}
\end{figure*}
\noindent\textbf{Targeted Query Processing.} This section analyzes the effectiveness of targeted query processing while running large data processing pipelines. We pick the ECG and ABP signals of several different dates from the real dataset with varying degrees of overlapping events between them to perform this analysis. Figure~\ref{fig:targetbench} shows the relative performance speedup of \streamer over Trill, measured with respect to the percentage of overlapping events in these data subsets. We observe that, as expected, the speedup is smaller when there is near perfect overlap, which is about $7\times$. The speedup starts to increase as the degree of overlap decreases, because \streamer can skip more and more redundant computations compared to Trill. For example, a day with $10\%$ overlap in ECG and ABP leads to about $38\times$ speedup over Trill, which is about $5\times$ higher than the base performance of \streamered.\\

\noindent\textbf{Window size}
In this section, we conduct a sensitivity study on \streamer to measure the effect of window size on its performance. Figure~\ref{fig:windowbench} shows the execution time of Trill and \streamer on the end-to-end benchmark over the synthetic dataset with window size varying from $1$ minute to $1$ hour. The results suggest that \streamer can maintain its performance benefits compared to Trill even on larger windows.

\subsection{Scalability}
Physiological datasets generally contain signals collected from thousands of patients, and the data processing pipelines usually process data from different patients separately. That means the data processing can be parallelized across multiple patients. \streamer takes advantage of this data-parallel nature of the physiological dataset to scale up the computation to both (i) multiple cores within a machine and (ii) multiple machines.

We evaluate the scalability of \streamer and compare it against Trill and numerical libraries on a single AWS m5a.8xlarge~\cite{aws} machine with $32$ cores and $128$ GB DRAM on the end-to-end benchmark using synthetic dataset. Figure~\ref{fig:multicorebench} shows total number of signal events processed per second against the number of parallel threads of data pipeline execution. We observe that \streamer provides up to $6.02\times$ better scalability than Trill and $1.90\times$ better than numerical libraries. This shows that \streamer can maintain its performance benefits on multi-core parallel data pipeline execution compared to Trill and provide better parallel processing capabilities than numerical library-based approach.

We also observe that Trill goes out of memory and crashes when we run experiments with more than $12$ parallel threads. \streamered, on the other hand, is much more memory-efficient and can scale up to $32$ parallel threads as the memory required for the intermediate results are preallocated are reused throughout query execution. Numerical library-based implementation is observed to scale up to $48$ threads; however, the performance gets saturated after $24$ threads and exhibits a peak performance that is $44\%$ lower than that of \streamered. 

We also measure the scalability of \streamer on a multi-machine setup using up to $16$ Amazon EC2 m5a.8xlarge~\cite{aws} machines with each running $12$, $24$ and $32$ parallel threads respectively for Trill, numerical libraries, and \streamered, since these thread counts are observed to provide the peak performance from multi-core experiment. Figure~\ref{fig:multimichbench} shows the throughput measured against the number of machines. On $16$ machines, \streamer can process $473.66$ million events per second which is $8.38\times$ higher than the peak performance of Trill and $1.73\times$ higher than that of numerical libraries. This shows that \streamer can maintain the performance benefits at large scale through efficient data-parallel processing.

\section{Related Work}
Several major solutions were proposed in the past to address large scale stream processing demands~\cite{spark, storm, dataflow, naiad}. Unfortunately, these solutions fail to be a good fit for physiological data processing due to the unique requirements of (i) programmability or (ii) performance (or sometimes even both). Below, we provide a detailed comparison of \streamer against key prior works based on these two aspects.\\

\noindent\textbf{Stream Processing Engines} Popular stream processing engines such as Apache Spark stream~\cite{structstream}, Storm~\cite{storm}, Flink~\cite{flink}, and Beam~\cite{dataflow} provide simple declarative programming interfaces for writing complex data processing pipelines. However, most of them fail to support several features essential for physiological data processing. For instance, Storm~\cite{storm} does not have any implicit notion of event time or windowing. Spark streaming, on the other hand, does not have support for millisecond precision event time as required for many signals in physiological data and lacks several useful temporal primitive operations that are necessary for writing queries on physiological data. Additionally, as we show in Section~\ref{sec:mot}, these solutions are primarily designed for distributed setup and showcase poor single machine performance.

Trill~\cite{trill} is the closest streaming engine we could find that provides rich support for temporal operations, high precision event time, and flexible windowing, as well as $1-2$ orders of magnitude higher single machine performance than distributed streaming engines. Unlike Trill, \streamer takes advantage of the end-to-end locality of the entire data pipeline using \emph{locality tracing}. Additionally, \streamer employs optimizations like \emph{static memory allocation} and \emph{targeted query processing} to minimize runtime memory allocation/deallocation overhead and pruning redundant computations. As we have shown in Section~\ref{sec:eval}, these optimizations make \streamer significantly faster than strong baselines like Trill.

A follow-up work called TrillDSP~\cite{trilldsp} extends Trill query language interface to support digital signal processing (DSP) operations like Fast Fourier Transform (FFT). Both \streamer and TrillDSP have similar motivations and they both deal with periodic data streams. However, TrillDSP is primarily focused on providing a programming interface that is suitable for people who are familiar with common signal processing libraries in R and MATLAB, simultaneously supporting stream processing capabilities of Trill without having to make expensive communications with these external libraries. \streamer proposes a more general solution which addresses both programmability and performance aspects of periodic stream processing. Moreover, the proposed query optimizations in our work, namely locality tracing, static memory allocation, and targeted query processing are unique contributions of \streamered.

Following the footsteps of Trill, there are two other streaming engines recently proposed, StreamBox~\cite{streambox} and StreamBox-HBM~\cite{streamboxhbm} that focus on improving the single machine performance. However, both these designs provide a very low-level and generic programming abstraction and lack a rich high-level temporal language support. Additionally, StreamBox-HBM was designed specifically for machines with high bandwidth memory (HBM), which are both very rare\footnote{In fact, Intel's Knight Landing architecture used in StreamBox-HBM is discontinued now~\cite{knl}.} and expensive. Compared to these two engines, \streamer provides a much simpler programming interface with high performance on commodity hardware.\\

\noindent\textbf{Numerical Libraries} There have been some recent studies~\cite{weld, split} to improve the performance of numerical library-based data processing pipelines. Weld~\cite{weld} proposed a compiler-based approach to optimize across disjoint libraries and functions with the help of an intermediate representation (IR). The followup work, called Split Annotations~\cite{split}, eliminates the need for an IR and re-implementation of library functions while potentially providing similar end-to-end performance benefits.

Even though these solutions can improve the end-to-end performance of numerical library-based data processing pipelines by some margin, the lack of temporal logic support and unified API specification still make such approaches less desirable in terms of ease of programming and maintainability. We believe high-level temporal query language provided by \streamer is more systematic and appropriate approach for doing data processing on physiological data.\\

\noindent\textbf{Time-series Databases} There are several databases specially designed to store and process time-series datasets efficiently. However, we find time-series databases like InfluxDB~\cite{influxdb} and KDB~\cite{kdb} to be a poor fit for physiological data processing. For instance, InfluxDB does not follow relational data model. Therefore, it does not have flexible temporal Join support, a common operation in the physiological data processing. Moreover, these databases only provide limited windowing support. KDB, on the other hand, requires the entire dataset to be in memory, which is not feasible as we are dealing with datasets of size up to $1$ TB. Additionally, KDB uses a highly esoteric language called \emph{q}~\cite{q} for writing queries in contrast to a more familiar SQL-like language.

We observe that time-series databases are sub-optimal choice from a performance perspective as well. InfluxDB introduces about $2.25\times$ overhead even on simple large-scale data read operation compared to stream processing engines. We expect even worse performance for more complicated operations.\\

\noindent\textbf{Stream Programming Languages} There has been prior work proposing programming language and compiler infrastructure for writing stream-based computation. Sisal~\cite{sisal} and StreamIt~\cite{streamit} are notable examples of a domain-specific language for writing streaming applications. However, these languages are not designed for temporal data processing and do not implicitly support event time. For example, temporal joining of two independent streams is not natively supported in both StreamIt and Sisal. Moreover, the optimizations in StreamIt are limited to DSP applications whereas the query language and the optimizations proposed in \streamer apply to generic temporal queries.

\section{Conclusion}
In this paper, we showcase the limitations of modern streaming engines and numerical libraries in building complex physiological data processing pipelines regarding their ease of programming, maintainability, and performance. We subsequently propose \streamered, which provides a simple and flexible temporal query language as the programming interface, and exploits the periodic nature of the physiological data to provide high performance. We propose three key optimizations in \streamered, namely, (i) locality tracing for improving end-to-end cache utilization of the data pipeline, (ii) memory footprint estimation for minimizing runtime memory allocation and deallocation overhead, and (iii) targeted query processing for pruning redundant computation. We conduct experiments and evaluations on real datasets and use cases, and demonstrate that \streamer outperforms state-of-the-art streaming engine Trill by as much as $7.5\times$ and numerical library-based approaches by as much as $3.2\times$ on the end-to-end data processing performance.

\section*{Acknowledgments}
We first thank our shepherd Abhishek Bhattacharjee and the anonymous reviewers for their valuable feedback and comments. We also like to thank members of the Laussen Lab, especially Mjaye Mazwi, Sebastian Goodfellow, Danny Eytan, Carson McLean, Sana Tonekaboni, Xi Huang, Aslesha Pokhrel, Sujay Nagaraj, and Bobby Greer for providing access to the physiological datasets at The Hospital for Sick Children and helping us build the benchmark suite. We further extend our gratitude to Jinliang Wei and the members of the EcoSystem lab, especially Shang Wang and Geoffrey Yu, for providing insightful comments and constructive feedback on the paper. This project was supported in part by the Canada Foundation for Innovation JELF grant, AWS machine learning research award, NSERC Discovery and CRD grants, and also a grant from Huawei.

\appendix
\section{Artifact Appendix}

\subsection{Abstract}
This artifact includes the source code of LifeStream and instructions to reproduce the main paper's key performance results. We identify the operations and end-to-end benchmark results described in Sections~\ref{sec:opbench} and \ref{sec:e2ebench} as the key results of the paper. The instructions provided are for reproducing the results shown in Figure~\ref{fig:opbench} and \ref{fig:e2ebench} on a synthetically generated data set. We include only the synthetic data set in the artifact since access to the real data set cannot be provided as it is proprietary to The Hospital for Sick Children, Toronto, Canada. The performance results produced on the synthetic data set should be an accurate estimate of the results on the real data set. Moreover, it should be noted that the results from this artifact are going to be better than the ones reported in the main paper as we have improved our results since the original submission.

The artifact can be executed over any runtime environment with .NET Core SDK 3.1 and Python3. However, the instructions provided in this appendix assumes a Ubuntu 20.04 machine. We also provide a docker file to automatically set up the runtime environment for running the artifact. We \textbf{recommend} using the docker environment for running the experiments.

\subsection{Artifact Checklist}
\begin{itemize}
  \item {\bf Algorithm:} Not applicable
  \item {\bf Program:} Custom benchmarks included in the artifact.
  \item {\bf Compilation:} C\# compiler with .NET Core 3.1 SDK
  \item {\bf Transformations:} No transformation tools required.
  \item {\bf Binary:} Source code and scripts included to build and run the binaries.
  \item {\bf Data set:} The synthetic data set used in the main paper is provided.
  \item {\bf Run-time environment:} .NET Core 3.1 SDK and Python3.
  \item {\bf Hardware:} A single general purpose CPU.
  \item {\bf Runtime state:} Not sensitive to runtime state.
  \item {\bf Execution:} Less than an hour to evaluate all the benchmarks.
  \item {\bf Metrics:} Data processing time in seconds.
  \item {\bf Output:} Plots similar to Figure~\ref{fig:opbench} and \ref{fig:e2ebench} in the main paper.
  \item {\bf Experiments:} Bash script and docker file are provided to run the benchmarks. Numerical variations in the results are negligible.
  \item {\bf How much disk space required (approximately)?:} Approximately $250$ MB.
  \item {\bf How much time is needed to prepare workflow (approximately)?:} Approximately $10$ minutes to setup the runtime environment.
  \item {\bf How much time is needed to complete experiments (approximately)?:} Approximately $30$ minutes.
  \item {\bf Publicly available?:} Yes
  \item {\bf Code licenses (if publicly available)?:} MIT License
  \item {\bf Data licenses (if publicly available)?:} Not applicable.
  \item {\bf Workflow framework used?:} No.
  \item {\bf Archival link:} \DOI
\end{itemize}

\subsection{Description}

\subsubsection{How to Access}
The artifact can be downloaded either from the GitHub link \github or from the DOI link \DOI.

\subsubsection{Hardware Dependencies}
LifeStream does not require any special hardware. A single general purpose CPU would be sufficient for running the artifact.

\subsubsection{Software Dependencies}
This artifact requires a runtime environment with .NET Core SDK 3.1 and Python3 installed. There are no restrictions on the choice of operating system. However, the instructions provided below assumes Ubuntu 20.04. Additionally, we also include a docker file to automatically setup the runtime environment.

\subsubsection{Benchmarks and Baselines}
This artifact uses the operations benchmark (Table~\ref{tab:opbench}) and end-to-end application benchmark (Figure~\ref{fig:pipeline}) used in the main paper. We include scripts to reproduce the results on Trill, numerical libraries (NumLib), and LifeStream for these benchmarks. We use execution time as the comparison metric for all the benchmarks.

\subsubsection{Data Sets}
In the main paper, we use two data sets: a synthetic data set and the real data set collected from The Hospital for Sick Children. The hospital regulations prevent us from sharing the real data set. Therefore, we only include the synthetic data set in the artifact which contains artificially generated $500$ Hz and $125$ Hz signal events. The results produced on the synthetic data set should give an accurate estimate of the results on real data set. For the operations benchmark, we use $1000$ minute $500$ Hz data set. End-to-end benchmark is evaluated on data set with both $500$ Hz and $125$ Hz signals with varying data sizes.

\subsection{Installation}
This section describes two ways for setting up the runtime environment; docker setup and manual setup. We recommend using docker environment for the experiments. First, download and unpack the artifact and go to the root directory.\\
For setting up the docker environment, 
\begin{enumerate}
    \item Build the docker image from the docker file by running the script \texttt{setup\_docker.sh}.
\end{enumerate}
For manually setting up the runtime environment,
\begin{enumerate}
    \item Go to \texttt{src} directory under the root directory.
    \item Install required packages by running \texttt{setup.sh} script.
\end{enumerate}

\subsection{Experiment Workflow}
Once the runtime environment has been properly setup, the following scripts can be used to run the benchmarks on Trill, numerical libraries (NumLib), and LifeStream.\\
For running in the docker setup, 
\begin{enumerate}
    \item Run \texttt{run\_docker.sh} script.
\end{enumerate}
For running in the manual setup,
\begin{enumerate}
    \item Go to \texttt{src} directory under the root directory.
    \item Run \texttt{run.sh} script.
\end{enumerate}

The above scripts run all the benchmarks one after the other and report the execution time for each benchmark on Trill, numerical libraries (NumLib), and LifeStream to the standard output as well as to respective files in the \texttt{src/results} directory.

\subsection{Evaluation and Expected Results}
Once the script has finished execution, they would have generated plots similar to Figure~\ref{fig:opbench} and \ref{fig:e2ebench}. The figures can be found in the \texttt{src/results} directory under the names \texttt{op.pdf} and \texttt{e2e.pdf}.

The plotted figures should show the performance comparison of LifeStream against Trill and numerical libraries (NumLib) on the synthetic data set. The results reported in the main paper, on the other hand, is based on the real data set, which we cannot share due to privacy concerns. Hence the evaluators should be comparing the relative performance of LifeStream against Trill and NumLib rather than the absolute numbers.

Compared to Trill, LifeStream is expected to perform $7-8\times$ better on end-to-end benchmark and $5-11\times$ better on operations benchmarks. Compared to NumLib, LifeStream should showcase performance improvement up to $\sim3\times$ on \emph{normalize}, \emph{resample}, and \emph{passfilter} benchmarks in operations benchmark. Results of LifeStream on \emph{fillconst} and \emph{fillmean} are expected to be worse than NumLib by up to $\sim2\times$ as reported in the main paper. It should be noted that the performance results of LifeStream on \emph{resample} and \emph{passfilter} benchmarks have been improved since the original submission as a result of several additional optimizations we incorporated. Finally, LifeStream should perform $2-3\times$ better than NumLib on the end-to-end benchmark.

\subsection{Experiment Customization}
To run a specific benchmark under manual setup, \texttt{run\_bench.sh} script under the \texttt{src} directory can be executed by passing the data set size (in seconds), execution engine, and the benchmark as follows.\\

\texttt{\$ ./run\_bench.sh <data size (sec)>} \\
\texttt{(trill | numlib | lifestream)}\\
\texttt{[ (normalize | passfilter | fillconst | fillmean}\\
\texttt{| resample | endtoend) ]}\\

\texttt{run\_bench.sh} script should report the execution time for the specific benchmark to the standard output.

To run a specific benchmark under the docker setup, open a shell to the docker container from the root directory using the following command before running the \texttt{run\_bench.sh} script.\\

\texttt{\$ docker run -it -v \$PWD/src/results:/root/results lifestream\_image /bin/bash}

\subsection{Methodology}

This artifact follows the following submission, reviewing and badging methodology:

\begin{itemize}[leftmargin=*]
\item \url{www.acm.org/publications/policies/artifact-review-badging}
\item \url{http://cTuning.org/ae/submission-20201122.html}
\item \url{http://cTuning.org/ae/reviewing-20201122.html}
\end{itemize}

\bibliographystyle{ACM-Reference-Format}

\end{document}